\documentclass[]{aa}  

\usepackage{color, colortbl}
\usepackage{xcolor}
\usepackage{amsmath}
\usepackage{graphicx}
\usepackage{txfonts}
\usepackage[colorlinks, urlcolor = blue, linkcolor = blue, citecolor=blue]{hyperref}
\usepackage{mathtools}
\usepackage{units}
\usepackage{subcaption}
\definecolor{Gray}{gray}{0.7}
\definecolor{lightGray}{gray}{0.85}
\definecolor{DarkGreen}{RGB}{00,124,00}
\usepackage{float}

\newcommand{\Msun}[0]{\,\mathrm{M_{\odot}}}
\newcommand{\Gaia}[0]{\textit{Gaia}}
\usepackage{soul}

\begin{document} 
\title{Expectations on the mass determination using astrometric microlensing by \Gaia{}}
\author{J. Kl\"{u}ter\inst{1},
	U. Bastian\inst{1},           
	J. Wambsganss\inst{1,}\inst{2}}

   \institute{ Zentrum f\"{u}r Astronomie der Universit\"{a}t Heidelberg, Astronomisches Rechen-Institut, M\"{o}nchhofstr. 12-14, 69120 Heidelberg, Germany\\
   \email{klueter@ari.uni-heidelberg.de}\and
   International Space Science Institute, Hallerstr. 6, 3012 Bern, Switzerland}
\authorrunning{J. Kl\"{u}ter et al.}
\titlerunning{Expectations on the mass determination using astrometric microlensing by \Gaia{}}

   \date{Received 05 November 2019/ Accepted 28 April 2020}
 
  \abstract
   {Astrometric gravitational microlensing can be used to determine the mass of a single star (the lens) with an accuracy of a few percent. To do so, precise measurements of the angular separations between lens and background star with an accuracy below \(1\,\text{milli-arcsecond}\) at different epochs are needed. Hence only the most accurate instruments can be used. 
However, since the timescale is in the order of months to years, the  astrometric deflection might be detected by \Gaia{}, even though  each star is only observed  on a low cadence.}
   {We want to show how accurately \Gaia{} can determine the mass of the lensing star.}
   {Using conservative assumptions based on the results of the second  \Gaia{} Data release, we simulated the individual  \Gaia{} measurements for 501 predicted astrometric microlensing events during the \Gaia{} era (2014.5 - 2026.5). For this purpose we use the astrometric parameters of \Gaia{} DR2, as well as an approximative mass based on the absolute G magnitude.
By fitting the motion of lens and source simultaneously we then reconstruct the 11 parameters of the lensing event. For lenses passing by multiple background sources, we also fit the motion of all background sources and the lens simultaneously.
Using a Monte-Carlo simulation we determine the achievable precision of the mass determination.
 } 
   {We find that \Gaia{} can detect the astrometric deflection for 114 events. Further, for 13 events \Gaia{} can determine the mass of the lens with a precision better than  \(15\%\) and for \(13+21 =  34\) events with a precision of \(30\%\) or better.}
{}


   \maketitle
%

\section{Introduction}
The mass is the most substantial parameter of a star. It defines its temperature, surface gravity and evolution. Currently, relations concerning stellar mass are based on binary stars, where a direct mass measurement is possible \citep{2010A&ARv..18...67T}. However, it is known that single stars evolve differently. Hence it is important to derive the masses of single stars directly. Apart from strongly model-dependent asteroseismology,  microlensing is the only usable tool. Either by observing astrometric microlensing events \citep{1991ApJ...371L..63P,1995AcA....45..345P} or by detecting finite source effects in photometric microlensing events and measuring the microlens parallax \citep{1992ApJ...392..442G}.
As sub-area of gravitational lensing, which was first described by Einstein's theory of general relativity \citep{1915SPAW...47..831E},  
microlensing describes the time-dependent positional deflection (astrometric microlensing) and magnification (photometric microlensing) of a background source by an intervening stellar mass. 
Up to now,  almost exclusively the photometric magnification was monitored and investigated by surveys such as  MOA\footnote{Microlensing Observations in Astrophysics} \citep{2001MNRAS.327..868B} or  OGLE\footnote{Optical Gravitational Lensing Experiment} \citep{2003AcA....53..291U}. 
By using the OGLE data in combination with simultaneous observations from the \textit{Spitzer} telescope, it was also possible to determine the mass of a few isolated objects  \citep[e.g.][] {2016ApJ...825...60Z,Chung_2017,Shvartzvald_2019,Zang_2020}.
Whereas the astrometric shift of the source was detected for the first time only recently
\citep[][]{2017Sci...356.1046S,2018MNRAS.480..236Z}. 
Especially \citet{2017Sci...356.1046S} showed the potential of
astrometric microlensing to measure the mass of a single star with a precision of a few per cent \citep{1995AcA....45..345P}. 
However, even though astrometric microlensing events are much rarer than photometric events, they can be \textit{predicted} from stars with known proper motions. The first systematic search for astrometric microlensing events was done by \citet{2000ApJ...539..241S}. 
Using the first Data Release of the \Gaia{} mission \citep{2016A&A...595A...1G}, \citet{2018MNRAS.478L..29M} predicted one event by a nearby white dwarf. 
Currently, the precise predictions 
(e.g. \citealt{2018A&A...615L..11K, 2018A&A...618A..44B}; \citealt{2018A&A...617A.135M}\footnote{\cite{2018A&A...617A.135M} searched for photometric micolensing events, however, for all of their predicted events a  measurable deflection is also expected.}; 
\citealt{2018AcA....68..183B,2018A&A...620A.175K})
make use of \Gaia{}'s second Data Release \citep{2018A&A...616A...1G} (hereafter \Gaia{} DR2)
or even combine \Gaia{} DR2 with external  catalogues \citep[e.g.][]{2018AcA....68..351N, 2019MNRAS.487L...7M}.

The timescales of astrometric microlensing events are typically longer than the timescales of photometric events.
 \citep[a few months instead of a few weeks;][]{2000ApJ...534..213D}. Hence, they might be detected and characterised by \Gaia{} alone, even though \Gaia{} observes each star only occasionally.
The \Gaia{} mission of the European Space Agency (ESA) is currently the most precise astrometric survey. Since mid-2014 \Gaia{} observes the full sky with an average of about 70 measurements within 5 years (nominal mission). 
\Gaia{} DR2 contains only summary results from the data analysis (J2015.5 position, proper motion, parallax etc.)  for its 1.6 billion stars, based on the first \(\sim\)2 years of observations.
However, with the fourth  data release (expected in 2024) and the final data release after the end of the extended mission, also the individual \Gaia{} measurements will be published. Using these measurements, it should be possible to determine the masses of individual stars using astrometric microlensing. 
This will lead to a better understanding of mass relations for main-sequence stars \citep{1991ApJ...371L..63P}.

In the present paper we show the potential of \Gaia{} to determine stellar masses using astrometric microlensing. We do so by simulating the individual measurements for 501 predicted microlensing events by 441 different stars.
We also show the potential of combining the data for multiple microlensing events caused by the same lens.

In Sect.\,\ref{chapter:microlensing} we describe astrometric microlensing. In Sect.\,\ref{chapter:gaia} we explain shortly the \Gaia{} mission and satellite, with a focus on important aspects for this paper. In Sect.\,\ref{chapter:Analysis} we  show our analysis, starting with the 
properties of the predicted events in\,\ref{section:Data}, the simulation of the \Gaia{} measurements in 
 \ref{section:Simulation}, the fitting procedure  \ref{section:reconstruction}, and  finally the statistical analysis in \ref{section:data_analysis}.
In Sect.\,\ref{chapter:Result}  we present the opportunities of direct stellar mass determinations by \Gaia{}.  Finally, we summarise the simulations and results and present our conclusions in Sect.\,\ref{chapter:conclusion}.


\section{Astrometric  Microlensing}
\label{chapter:microlensing}

\begin{figure}
\includegraphics[width=9cm]{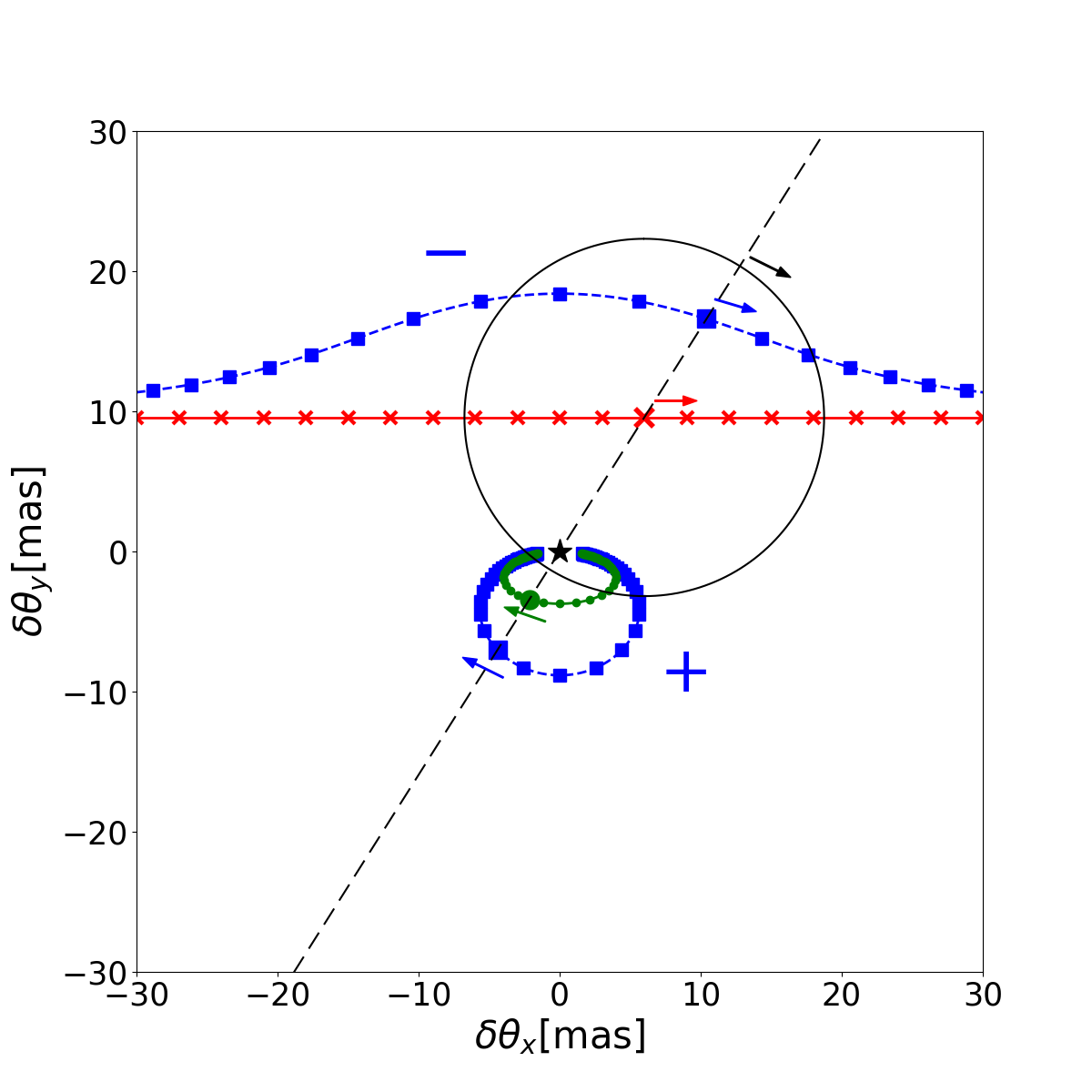}
\caption{The astrometric shift for an event with an Einstein radius of 
\(\theta_{E} = 12.75 \mathrm{mas}\) (black circle) and an impact parameter of \(u = 0.75\). 
While the lens (red) passes a background star (black star, fixed in origin) two images  (blue dashed, major image + and minor image - ) of the 
source are created due to gravitational lensing. This leads to a shift of the centre of light, shown in 
green. The straight long-dashed black line connects the positions of the images for one epoch. While the lens is moving in the direction of the 
red arrow, all other images are moving according to their individual arrows. The red, blue and 
green dots correspond to different epochs with fixed time steps \citep[after][]{2011A&A...536A..50P}.}
\label{figure:shift}
\end{figure}

The change of the centre of light  of the background star (``source'') due to the gravitational deflection of a passing foreground star (``lens'') is called astrometric microlensing. This is shown in Figure\,\ref{figure:shift}. While the lens (red line) is passing the 
source (black star in the origin), two images of the source are created (blue lines): a bright major image \((+)\) close to the unlensed position, and a faint minor image\,\((-)\) close to the lens.  In case of a perfect alignment, both images merge to an Einstein ring, with a radius of \citep{1924AN....221..329C,1936Sci....84..506E,1986ApJ...301..503P}:
\begin{equation}
\theta_{E} = \sqrt{\frac{4GM_{L}}{c^{2}}   \frac{D_{S}-D_{L}}{D_{S}\cdot D_{L}}} = 2.854\,\mathrm{mas} \sqrt{\frac{M_{L}}{\Msun{}}\cdot\frac{\varpi_{L}-\varpi_{S}}{1\,\mathrm{mas}}},
\label{equation:Einsteinradius}
\end{equation}
where \(M_{L}\) is the mass of the lens and \(D_{L}\), \(D_{S}\) are the distances of the lens and source from the observer, and \(\varpi_{L}\),\(\varpi_{S}\) are the parallaxes of lens and source, respectively. \(G\) is the gravitational constant, and \(c\) the speed of light. 
For a solar-type star at a distance of about \(1\,\mathrm{kiloparsec}\) the Einstein radius is of the order of a few milli-arc-seconds (\(\mathrm{mas}\)).
The Einstein radius defines an angular scale of the microlensing event.
Using the unlensed scaled angular separation on the sky \(\boldsymbol{u} = \boldsymbol{\Delta\phi}/\theta_{E}\), where \(\boldsymbol{\Delta\phi}\) is the two-dimensional unlensed angular separation, the position of the two lensed images can be expressed as a function of \(\boldsymbol{u}\), by \citep{1996ARA&A..34..419P}
\begin{equation}
\boldsymbol{\theta_{\pm}} = \frac{ u \pm \sqrt{u^{2}+4}}{2} \cdot \frac{\boldsymbol{u}}{u} \cdot{\theta_{E}},
\label{equation:position_lens}
\end{equation}
with \(u = \lvert \boldsymbol{u} \rvert \).

For the unresolved case, only the centre of light of both images (green line) can be observed. 
This can be expressed by \citep{1995A&A...294..287H,1995AJ....110.1427M,1995ApJ...453...37W}: 
\begin{equation}
\boldsymbol{\theta_c}= \frac{A_{+}\boldsymbol{\theta_{+}} +A_{-}\boldsymbol{\theta_{-}}}{A_{+}+A_{-}} =\frac{u^{2}+3}{u^{2}+2}\boldsymbol{u}\cdot{\theta_{E}},
\end{equation}
where \(A_{\pm}\) are the magnifications of the two images given by \citep{1986ApJ...301..503P}
\begin{equation}
A_{\pm} =  \frac{ u^{2}+2}{2u\sqrt{u^{2}+4}}\pm 0.5 .
\label{equation:magnification}
\end{equation}
The corresponding angular shift is given by
\begin{equation}
\delta\boldsymbol{\theta_{c}} = \frac{\boldsymbol{u}}{u^{2}+2} \cdot{\theta_{E}}.
\label{equation:shift}
\end{equation}
The measurable deflection can be further reduced due to luminous-lens effects. However, in the following, we  consider the resolved case, where luminous-lens effects can be ignored. 
Due to equations (\ref{equation:position_lens}) and (\ref{equation:magnification}), the influence of the minor image can only be observed when the impact parameter is in the same order of magnitude as or smaller than the Einstein radius. Therefore the minor image is only hardly resolvable and so far was resolved only once \citep{Dong_2019}. 
Hence, the observable for the resolved case is only the shift of the position of the  major image. This can be expressed by
\begin{equation}
\label{equation:microlensing_term}
\delta\boldsymbol{\theta_{+}} = \frac{  \sqrt{(u^{2}+4)} - u}{2} \cdot \frac{\boldsymbol{u}}{u} \cdot{\theta_{E}}.
\end{equation}
For large impact parameters \(u\gg 5\) this can be approximated as \citep{2000ApJ...534..213D}
\begin{equation}
\delta\theta_{+} \simeq \frac{\theta_{E}}{u} = \frac{\theta_{E}^{2}}{ \lvert\boldsymbol{\Delta \phi} \rvert} \propto \frac{M_{L}}{ \lvert\boldsymbol{\Delta\phi} \rvert}
\label{equation:approx_shift}. 
\end{equation}
which is  proportional to the mass of the lens.
Nevertheless equation\,(\ref{equation:shift}) is also a good approximation for the shift of the  major image whenever \(u > 5\), since then the second image is negligibly faint. This is always the case in the present study.

\section{\Gaia{} satellite}
\label{chapter:gaia} 
The \Gaia{} satellite is a space telescope of the European Space Agency (ESA) which was launched in December 2013. It is located at the Earth-Sun  Lagrange point L2, where it orbits the sun at roughly \(1\%\) larger distance than the earth. In mid 2014 \Gaia{} started to observe the whole sky on a regular basis defined by a nominal (pre-defined) scanning law.

\subsection{Scanning law}
The position and orientation of \Gaia{}  is defined by various periodic motions.  First, it rotates with a period of 6 hours around itself. 
Further, \Gaia{}'s spin axis is inclined by 45 degrees to the sun, with a precession frequency of one turn around the sun every 63 days. 
Finally, \Gaia{} is not fixed at L2 but moving on a \(100\,000\,\mathrm{km}\) Lissajous-type orbit around L2.
The orbit of \Gaia{} and the inclination is chosen such that the overall coverage of the sky is quite uniform with 
about 70 observations per star during the nominal 5-year mission \citep[2014.5 to 2019.5]{2016A&A...595A...1G} in different scan angles. However, certain parts of the sky are inevitably observed more often. 
Consequently, \Gaia{} cannot be pointed on a certain target at a given time. 
We use the \Gaia{} observation forecast tool (GOST)\footnote{\Gaia{} observation forecast tool, \\ \url{https://gaia.esac.esa.int/gost/}} to get the information when a target is 
inside the field of view of \Gaia{}, and the current scan direction of \Gaia{} at each of those times. 
GOST also lists the CCD row, which can be translated into eight or nine CCD observations.
For more details on the scanning law see \citet{2016A&A...595A...1G} or the \Gaia{} Data Release Documentation\footnote{\Gaia{} Data Release Documentation - The scanning law in theory \\ \url{https://gea.esac.esa.int/archive/documentation/GDR2/Introduction/chap_cu0int/cu0int_sec_mission/cu0int_ssec_scanning_law.html}}

\subsection{Focal plane and readout window} 
\begin{figure}
\includegraphics[width = 9cm]{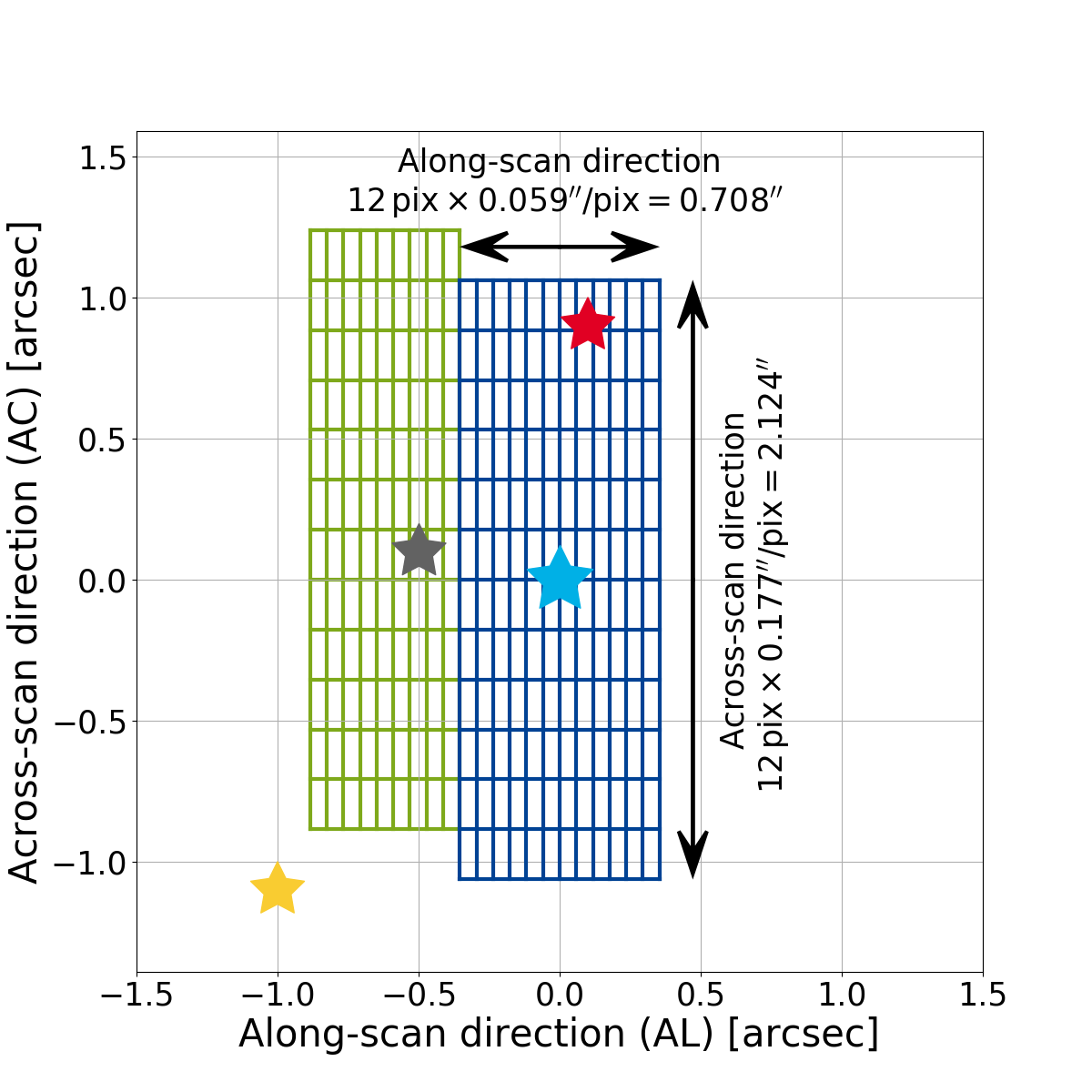}
\caption{Illustration of the readout windows. For the brightest source 
(blue big star) \Gaia{} assigns the full window (blue grid) of 
\(12 \times 12\,  \mathrm{pixels}\) When a second source is within this 
window (e.g. red star) we assume that this star is not observed by \Gaia{}. If 
the brightness of both stars are similar (\(\Delta G < 1\,mag\)) we neglect also the 
brighter source. For a second source close by but outside of the readout 
window (e.g. grey star) \Gaia{} assigns a truncated readout window 
(green grid). We assume that this star can be observed, and the precision in 
along scan direction is the same as for the full readout window. For more distant sources (e.g. yellow star) \Gaia{} assigns a full window.}
\label{figure:window}
\end{figure}

\Gaia{} is equipped with two separate telescopes with rectangular primary mirrors, pointing on two  fields of view, separated by \(106.5^{\circ}\). This results in two observations only a few hours apart with the same scanning direction.
The light of the two fields of view is focused on one common focal plane which is 
equipped with 106 CCDs arranged in 7 rows. The majority of the CCDs (62) are used for the 
astrometric field. While \Gaia{} rotates, the source first passes a sky mapper, 
which can distinguish between both fields of view. Afterwards, it passes nine CCDs of the 
astrometric field (or eight for the middle row).  The astrometric field is devoted to position measurements, 
providing the astrometric parameters, and also the G-band photometry. For 
our simulations, we stack the data of these eight or nine CCDs into one measurement.  Finally, the source passes a red and blue photometer, plus a radial-velocity spectrometer  \citep{2016A&A...595A...1G}.
In order to reduce the volume of data, only small "windows" around detected sources 
are read out and transmitted to ground. For faint sources  \((G >13\,\mathrm{mag})\)  these windows 
are  \(12 \times 12\,\mathrm{pixels}\) 
(\(\text{along-scan}\times \text{across-scan}\)).
This corresponds to \(708\,\mathrm{mas}\, \times\, 2124 \, \mathrm{mas}\), due to a 1:3 pixel-size ratio. 
These data are stacked by the onboard processing of \Gaia{} in across-scan direction into a one-dimensional strip, which is then transmitted to Earth. 
For bright sources \((G <13\,\mathrm{mag})\) larger windows (\(18\,\times\,12\,\mathrm{pixel}\)) are read out. These data are transferred as 2D images
\citep{2016A&A...595A...7C}. 
When two sources with overlapping readout windows (e.g. Fig.\,\ref{figure:window}:  blue and grey stars) are detected, Gaia{}'s onboard processing assigns the full window (blue grid) only to one of the sources (usually the brighter source). 
For the second source \Gaia{} assigns only a truncated window (green grid).  For \Gaia{} DR2 these truncated windows are not processed\footnote{\Gaia{} Data Release Documentation - Datamodel description   \\ \url{https://gea.esac.esa.int/archive/documentation/GDR2/Gaia_archive/chap_datamodel/sec_dm_main_tables/ssec_dm_gaia_source.html}}.
For more details on the focal plane and readout scheme  see \cite{2016A&A...595A...1G})
 
\subsection{Along-scan precision}

\begin{figure}
\includegraphics[width = 9 cm]{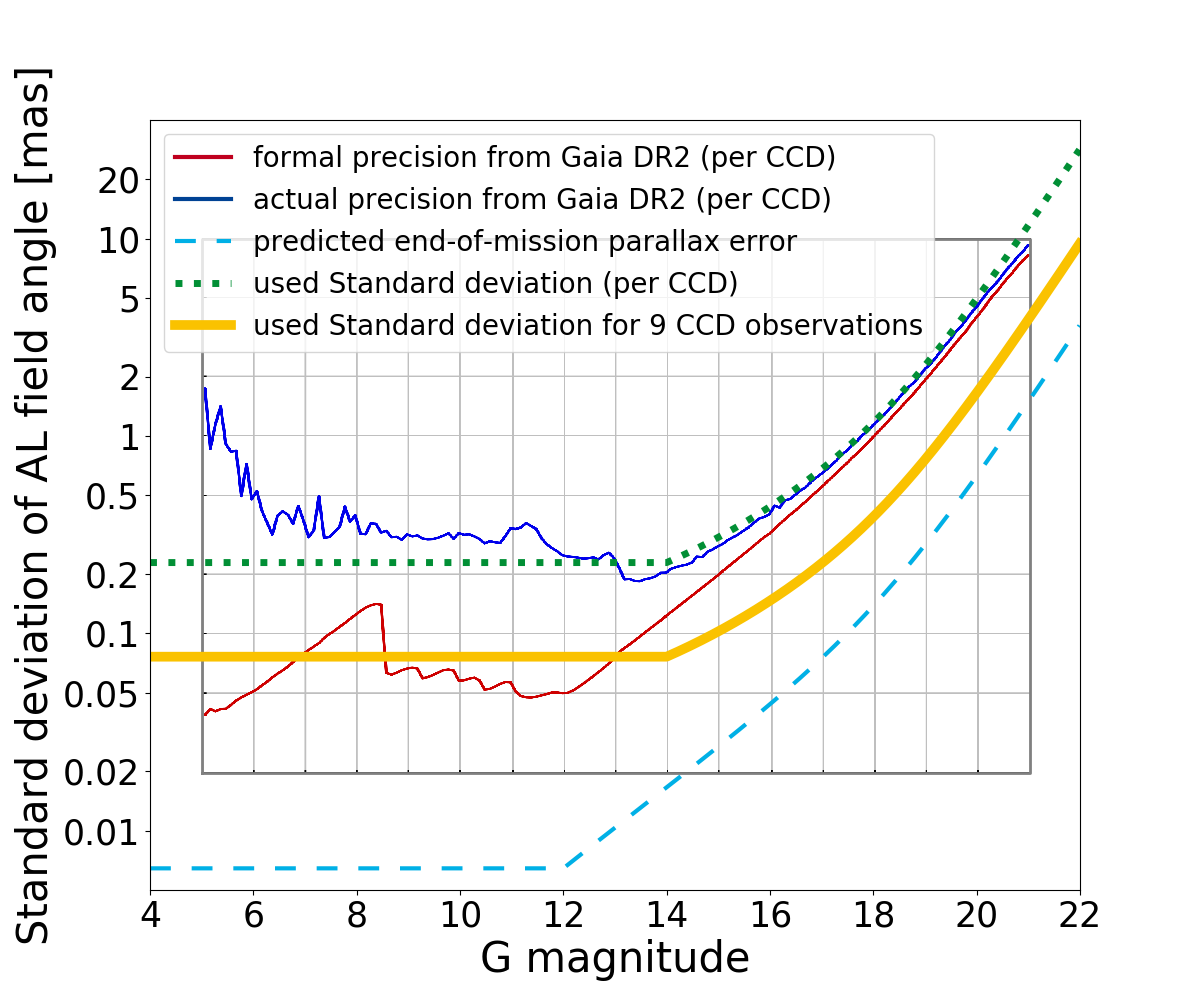}
\caption{Precision in along-scan direction as function of G magnitude.
The red line indicates the expected formal precision from \Gaia{} DR2 for one CCD observation. The blue solid line is the actually achieved precision \citep{2018A&A...616A...2L}. The light blue dashed line shows the relation for the end-of-mission parallax error \citep{2016A&A...595A...1G}, and the green dotted line shows the adopted relation for the precision per CCD observation  for the present study. The adopted precision for 9 CDD observations is shown as thick yellow curve.
The inlay (red and blue curve) is taken from \citet{2018A&A...616A...2L}, Fig.\,9.}
\label{figure:sig_gmag}
\end{figure}

Published information about the precision and accuracy of \Gaia{} mostly 
refers to the end-of-mission standard errors, 
which result from a combination of all individual measurements, and also consider the different scanning directions. 
\Gaia{} DR1 provides an analytical formula to estimate this precision as a 
function of G magnitude and V-I colour \citep{2016A&A...595A...1G}.  
However, we are interested in the precision of one single field-of-view 
scan (i.e. the combination 
of the nine or eight CCD measurements in the astrometric field). The red line in 
Figure\,\ref{figure:sig_gmag} shows the formal 
precision in along-scan direction for one CCD \citep{2018A&A...616A...2L}. The precision is 
mainly dominated by photon noise. Due to different readout gates, the number of photons is roughly 
constant for sources brighter than 
\(G = 12\,\mathrm{mag}\).
The blue line in Figure\,\ref{figure:sig_gmag} shows the actual scatter of the post-fit 
residuals, and the difference represents the combination of all unmodeled 
errors. For more details on the precision see \citet{2018A&A...616A...2L}.

\section[title]{Simulation of  \Gaia{} Measurements and Mass reconstruction\footnote{The Python-based code for our simulation is made publicly available \url{https://github.com/jkluter/MLG}}}

\label{chapter:Analysis}
\begin{figure}
\includegraphics[width = 9cm]{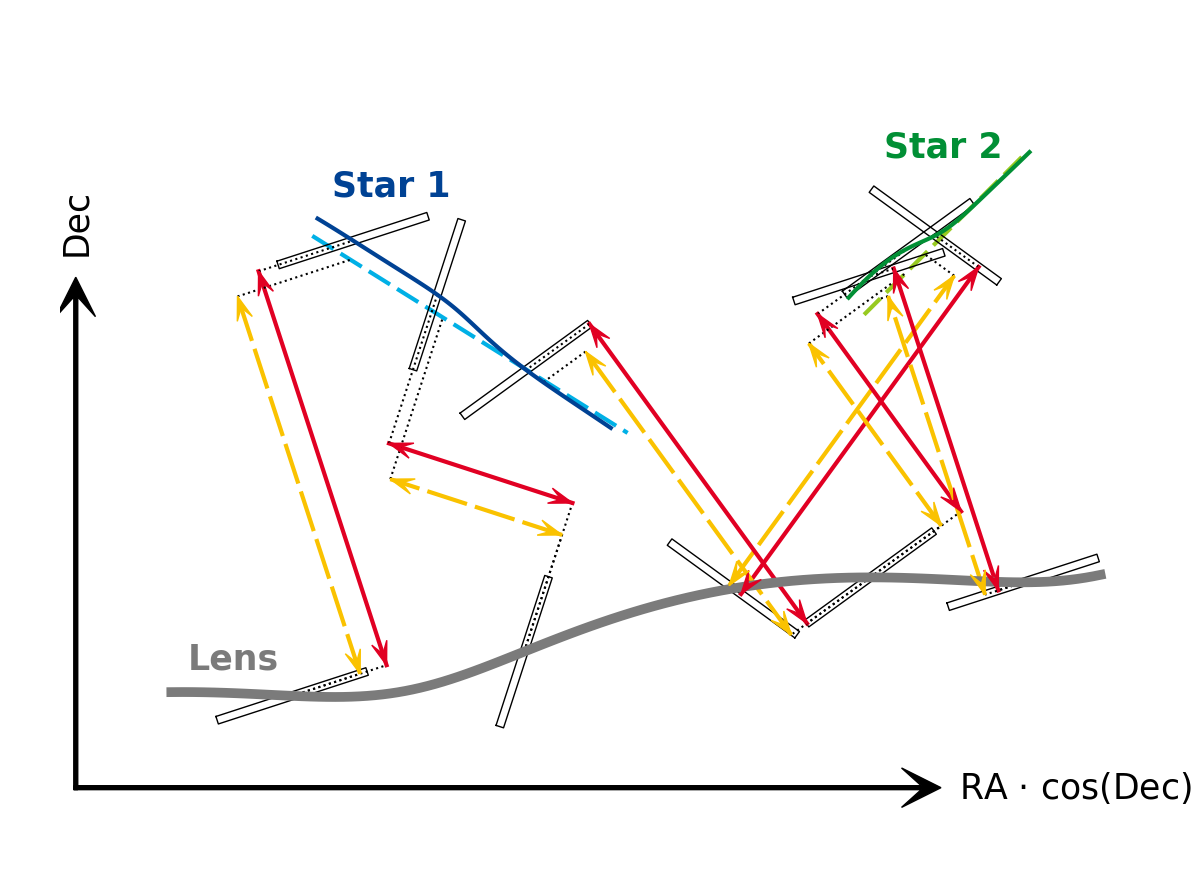}

\caption{Illustration of our simulation. While the lens (thick grey line) passes 
the background star\,1  (dashed blue line) the observed position of the background star is 
slightly shifted due to microlensing (solid blue line). The \Gaia{} measurements 
are indicated as black rectangles, where the precision in along-scan direction is much
better than the precision in across-scan direction.
The red arrows 
indicate the along-scan separation including microlensing, and the yellow 
dashed arrows show the along-scan separation without microlensing. The 
difference between both sources shows the astrometric microlensing 
signal.  Due to the different scaning direction, an observation close to the maximal deflection of the microlensing event does not necessarily have the largest signal.  
A further background star\,2 (green) can improve the result.}
\label{figure:Illustration}
\end{figure}

The basic structure of the simulated dataset is illustrated in Figure~\ref{figure:Illustration}.
We start with the selection of lens and source stars (Subsect.\,\ref{section:Data}). Afterwards, we calculate the observed positions using the \Gaia{} DR2 
positions, proper motions, and parallaxes, as well as an assumed mass (Subsect.\,\ref{subsection:Astrometry}; Fig.\,\ref{figure:Illustration} solid grey and solid blue lines). We then determine the one-dimensional \Gaia{} measurements (Subsect.\,\ref{subsection:resolution} and \ref{subsection:measurement}; Fig.\,\ref{figure:Illustration} black rectangles).
Finally, we fit the simulated data (Subsect.\,\ref{section:reconstruction}) using only the residuals in along-scan direction. 
We repeat these steps to estimate the expected uncertainties of the mass determination via a Monte-Carlo approach (Subsect.\,\ref{section:data_analysis}).

\subsection{Data input}
\label{section:Data}

We simulate 501 events predicted by  \citet{2018A&A...620A.175K} with an 
epoch of the closest approach between 2013.5 and 2026.5. 
We also include the events outside the most extended mission of \Gaia{} (ending in 2024.5), 
since it is possible to determine the mass only from the tail of an event (e.g. event \#62 - \#65 in Tab.~\ref{table:single2}), or from 
using both \Gaia{} measurements and additional observations. 
The sample is naturally divided into two categories: events where the 
motion of the background source is known, and events where the 
motion of the background source is unknown. A missing proper motion in DR2 will not automatically mean 
that \Gaia{} cannot measure the motion of the background source. The data for 
\Gaia{} DR2 are derived only from a 2-year baseline. 
With the 5-year baseline for the nominal mission ended in mid-2019 and 
also with the potential extended 10-year baseline, \Gaia{} is expected to 
provide proper motions and parallaxes also for 
some of those sources.
In order to deal with the unknown proper motions and parallaxes we use randomly 
selected values
from a normal distribution with 
means of \(5\,\mathrm{mas/yr}\) and \(2\,\mathrm{mas}\),  respectively and 
standard deviations of   \(3\,\mathrm{mas/yr}\) and \(1\,\mathrm{mas}\), 
respectively, and using a uniform distribution for the direction of the  proper 
motion. For the parallaxes, we only use the positive part of the distribution. 
Both distributions roughly reflect the sample of all potential background 
stars in  \citet{2018A&A...620A.175K}.

\subsubsection*{Multiple background sources}
Within 10 years, some of our lensing stars pass close enough to multiple background stars, thus causing several measurable astrometric effects. 
As an extreme case, the light deflection by Proxima Centauri causes a measurable shift (larger than 
\(0.1\,\mathrm{mas}\)) on 18 background stars. This is  due to the star's large Einstein radius, its high proper motion 
and the dense background. Since those events are physically connected, we simulate and 
fit  the motion of the lens and multiple background sources simultaneously (see Fig.\,\ref{figure:Illustration}, Star 2).
We also compare three different scenarios: A first one where we use all background sources, a 
second one  where we only select those with known proper motion, and a  third one where we only select 
those with a precision in along-scan direction better than \(0.5\,\mathrm{mas}\) per field of view transit (assuming 9 CCD observations). The latter limit corresponds roughly to sources brighter 
than \(G \simeq 18.5\,\mathrm{mag}\).

\subsection{Simulation of \Gaia{} Data}
\label{section:Simulation}

We expect that \Gaia{} DR4 and the full release of the extended mission will provide for each single CCD observation the position and uncertainty in along scan direction, in combination with the observation epochs. These data are simulated as a basis for the present study. 
We thereby assume that all variations and systematic effects caused by the satellite itself are corrected beforehand. 
However, since we are only interested in relative astrometry, measuring the astrometric deflection is not affected by most of the systematics as for example the slightly negative parallax zero-point \citep{2018A&A...616A...9L}. 
We also do not simulate all CCD measurements separately, but rather a mean measurement of all eight or nine  CCD measurements during a field of view transit. 
In addition to the astrometric measurements,  \Gaia{} DR4 will also publish the scan angle and the barycentric location of the \Gaia{} satellite.

We find that our results  strongly depend on the temporal distribution of measurements and their scan directions. Therefore we use for each event 
predefined epochs and scan angles, provided by the GOST online tool. This tool only lists the times and angles when a certain area is passing the field of view of \Gaia{}. 
However, it is not guaranteed that a measurement is actually taken and transmitted to Earth. We assume that  for each transit  \Gaia{} measures the position of the
background source and lens simultaneously (if resolvable), with a certain probability for missing data points and clipped outliers.   

To implement the parallax effect for the simulated measurements we assume that the position of the \Gaia{} satellite is exactly at a 1\% larger 
distance to the Sun than the Earth. Compared to a strict treatment of the actual \Gaia{} orbit, we do not expect any differences in the results, since first, \Gaia{}'s distance from this point (roughly L2) is very small compared to the distance to the Sun, and second, we consistently use 1.01 times the earth orbit for the simulation and for the fitting routine. 
The simulation of the astrometric \Gaia{} measurements is described in the following subsections.

\subsubsection{Astrometry}
\label{subsection:Astrometry}
Using the \Gaia{} DR2 positions (\(\alpha_{0},\delta_{0}\)), proper motions 
(\(\mu_{\alpha*,0},\mu_{\delta,0}\))  and parallaxes (\(\varpi_{0}\)) we calculate the unlensed 
positions of lens and background source seen by \Gaia{} as a function of time (see Fig.\,\ref{figure:Illustration} solid grey line and dashed blue line), using the following equation: 
\begin{equation} 
 \begin{pmatrix} \alpha \\ \delta \end{pmatrix} =  \begin{pmatrix} \alpha_{0} \\ \delta_{0} \end{pmatrix}+ (t-t_{0})
\begin{pmatrix} \mu_{\alpha*,0}/\cos \delta_{0} \\ \mu_{\delta,0} \end{pmatrix}+1.01\cdot \varpi_{0} \cdot \vec{J}^{-1}_{ \ominus}  \vec{E(t)},
\label{equation:motion}
\end{equation}
where \(\vec{E(t)}\) is the barycentric position of the Earth,  in cartesian coordinates, 
in astronomical units and 

\begin{equation}
 \vec{J}^{-1}_{ \ominus} = \begin{pmatrix} \sin\alpha_{0}/\cos\delta_{0}&-\cos\alpha_{0}/\cos\delta_{0}&0\\ \cos\alpha_{0}\sin\delta_{0}   &\sin\alpha_{0}\sin\delta_{0}&-\cos\delta_{0}
\end{pmatrix}
\end{equation}
is the inverse Jacobian matrix for the transformation into a spherical coordinate system, 
evaluated at the lens position. 

We then calculate the observed position of the source (see Fig.\,\ref{figure:Illustration} solid blue line) by adding the microlensing term (Eq.\,(\ref{equation:microlensing_term})). 
Here we assume that all our measurements are in the resolved case. That means, \Gaia{} observes the position of the  major image of the source, and the measurement of the lens position is not affected by the  minor image of the source.
For this case the exact equation is: 
\begin{equation}
\label{equation:microlensing_excact}
\begin{pmatrix} \alpha_{obs} \\ \delta_{obs} \end{pmatrix}  =  \begin{pmatrix} \alpha \\ \delta \end{pmatrix} + \begin{pmatrix} \Delta \alpha \\ \Delta\delta \end{pmatrix} \cdot \left(\sqrt{0.25+\frac{\theta_{E}^{2}}{\Delta\phi^{2}}}-0.5\right)
\end{equation}
\\
where \(\Delta\phi = \sqrt{(\Delta\alpha\cos \delta)^{2}+(\Delta\delta)^{2}} \) is the unlensed angular separation between lens and source and 
\((\Delta\alpha, \Delta\delta) = (\alpha_{source}-\alpha_{lens}, \delta_{source}-\delta_{lens})\) 
are the differences in right ascension and declination, respectively.
However, this equation shows an unstable behaviour in the fitting process, caused by the square root. This results in a time-consuming fit process. 
To overcome this problem we use the shift of the centre of light as approximation for the shift of the brightest image. This approximation is used for both the simulation of the data and the fitting procedure:
\begin{equation}
\label{equation:microlensing}
\begin{pmatrix} \alpha_{obs} \\ \delta_{obs} \end{pmatrix}  =  \begin{pmatrix} \alpha \\ \delta \end{pmatrix} + \begin{pmatrix} \Delta \alpha \\ \Delta\delta \end{pmatrix} \cdot \frac{\theta_{E}^{2}}{\left(\Delta\phi^{2}+2 \theta_{E}^{2} \right)}, 
\end{equation}
The differences between equations\,(\ref{equation:microlensing_excact}) and   (\ref{equation:microlensing}) are by at least a factor 10  smaller than the measurements errors (for most of the events even by a factor 100 or more). Further, using this approximation we underestimate the microlensing effect, thus being on a conservative track for the estimation of mass determination efficiency.

We do not include any orbital motion in this analysis even though SIMBAD listed some of the lenses (e.g. 75 Cnc) as binary stars. However, from an inspection of their orbital parameters \citep[e.g. periods of a few days]{2004A&A...424..727P}  we expect that this effect influences our result only slightly. The inclusion of orbital motion would only be meaningful if a good prior would be available. This might come with \Gaia{} DR3 (expected for end of 2021).

\subsubsection{Resolution}
\label{subsection:resolution}
Due to the on-board readout process and the on-ground data processing the resolution of \Gaia{} is not limited by its point-spread function, but limited by the size of the readout windows\footnote{This is a conservative assumption. It is true for Gaia Data Releases DR1,2,3, but for DR4 and DR5 there are efforts under way to essentially get down to the optical resolution.}. 
Using the apparent position and G magnitude of lens and source we investigate for all given epochs if \Gaia{} can resolve both stars
or if \Gaia{} can only measure the brightest of both (mostly the lens, see Fig.\,\ref{figure:window}). 
We therefore calculate the separation in along-scan and across-scan direction, as 
\begin{equation}
\begin{aligned}
\Delta\phi_{AL} &= \mid\sin \Theta \cdot \Delta\alpha\cos \delta +    \cos \Theta \cdot \Delta\delta\mid \\
\Delta\phi_{AC} &= \mid - \cos \Theta \cdot \Delta\alpha\cos \delta +   \sin \Theta \cdot \Delta\delta\mid,
\end {aligned}
\end{equation}
where \(\Theta\) is the position angle of scan direction counting from North towards East. 
When the fainter star is outside of the read out window of the brighter star,  
that means the separation in along-scan direction is larger than \(354\,\mathrm{mas}\)  or the separation in across scan direction is larger \(1062\,\mathrm{mas}\), we assume that \Gaia{} measures the positions of both sources. Otherwise we assume that 
only the position of the brightest star is measured, unless both sources have a similar brightness (\(\Delta G < 1\,\mathrm{mag}\)). In that case, we exclude the measurements of both stars.

\subsubsection{Measurement errors}
\label{subsection:measurement}

In order to derive a relation for the uncertainty in along-scan direction as function of the G magnitude, we start with the equation for the end-of-mission parallax standard error, where we ignore the additional 
colour term \citep[see Fig.\,\ref{figure:sig_gmag}, blue dashed line]{2016A&A...595A...1G}:   
\begin{equation}
\label{equation:sigma_parallax}
\sigma_{\varpi} = \sqrt{-1.631 + 680.766 \cdot z + 32.732 \cdot z^{2}} \,\mathrm{\mu as}
\end{equation}
with 
\begin{equation}
\label{equation:z_original}
z  = 10^{(0.4\,(\max(G,\,12) - 15))} 
\end{equation}
We then adjust this relation in order to describe the actual precision in along-scan direction per CCD shown in \citet{2018A&A...616A...2L} (Fig.\,\ref{figure:sig_gmag}, blue line) by multiplying by a factor of  \(7.75\) and adding an offset of \(100\,\mathrm{\mu as}\). 
And we adjust \(z\) (Eq.\,(\ref{equation:z_original})) to be constant for \(G < 14\,\mathrm{mag}\)
(Fig.\,\ref{figure:sig_gmag}, green dotted line).  These adjustments are done heuristically.
We note that we overestimate the precision for bright 
sources, however most of the background sources, which carry the astrometric microlensing signal, are fainter than \(G = 13\,\mathrm{mag}\).  For those sources the assumed precision is slightly worse compared to the actually achieved precision for \Gaia{} DR2. 
Finally we assume that during each field-of-view transit all nine (or eight) CCD observations are useable. Hence, we divide the CCD precision by \(\sqrt{N_{CCD}} = 3\, (\text{or } 2.828)\) to determine the standard error in along-scan direction per field-of-view transit: 
\begin{equation}
\sigma_{AL} =\frac{\left(\sqrt{-1.631 + 680.766 \cdot \tilde{z}  + 32.732 \cdot \tilde{z} ^{2}}\cdot 7.75+100\right)}{\sqrt{N_{CCD}}}\, \mathrm{\mu as}
\end{equation}
with 
\begin{equation}
\tilde{z}  = 10^{(0.4\,(\max(G,\,14) - 15))} 
\end{equation}
In across-scan direction we assume a precision of \(\sigma_{AC} = 1''\). This is only used as rough estimate for the simulation, since  only the along-scan component is used in the fitting routine. 

For each star and each field-of-view transit we pick a value from a 2D Gaussian distribution with \(\sigma_{AL}\) and \(\sigma_{AC}\) in along-scan and across-scan direction, respectively, as positional measurement. 

Finally, the data of all resolved measurements are forwarded to the fitting routine.  
These contain the positional measurements \((\alpha,\,\delta)\), the standard error in along-scan direction\((\sigma_{AL})\), the epoch of the  observation (\(t\)), the current scanning direction (\(\Theta\)), as well as an identifier for the corresponding star (i.e. if the measurement corresponds to the lens or source star).

\subsection{Mass reconstruction}
\label{section:reconstruction}
To reconstruct the mass of the lens we fit equation\,(\ref{equation:microlensing}) (including the dependencies of Eq.\,(\ref{equation:Einsteinradius}) and (\ref{equation:motion})) to the data of the lens and the source simultaneously. 
For this we use a weighted-least-squares method.
Since \Gaia{} only measures precisely in along scan direction (see Fig.\,\ref{figure:Illustration} black boxes), we compute the weighted  residuals \(r\) as follow:
\begin{equation}
r =\frac{\sin\Theta\,(\alpha_{model} - \alpha_{obs})\cdot\cos{\delta}+  \cos\Theta\, (\delta_{model} - \delta_{obs})}
{\sigma_{AL}},
\label{equation:residuen}
\end{equation} while ignoring the across-scan component.
The open parameters of this equation are the mass of the lens as well as the \(5\) astrometric parameters of the lens and each source. This adds up to 11 fitted parameters for a single event, and \(5\times n+6\) fitted parameters for the case of n background sources (e.g. \(5\times18+6=96\) parameters for the case of 18 background sources of Proxima Centauri)

The used least-squares method is a Trust-Region-Reflective algorithm \citep{S1064827595289108}, which we also provide with the analytic form of the Jacobian matrix of equation\,(\ref{equation:residuen})
(including all inner dependencies from Eq.\,(\ref{equation:Einsteinradius}), (\ref{equation:motion}) and (\ref{equation:microlensing})).
We do not exclude negative masses, since, due to the noise, there is a non-zero probability that the determined mass will be below zero.
As initial guess, we use the first data point of each star as position, along with zero parallax, zero proper motion, as well as a mass of \(M = 0.5 \Msun{}\). One could use the motion without microlensing to analytically calculate an initial guess, however, we found that this neither improves the results nor reduces the computing time significantly. 

\subsection{Data analysis}
\label{section:data_analysis}
\begin{figure}
\begin{subfigure}{.25\textwidth}
\centering
\includegraphics[width = 4.5cm]{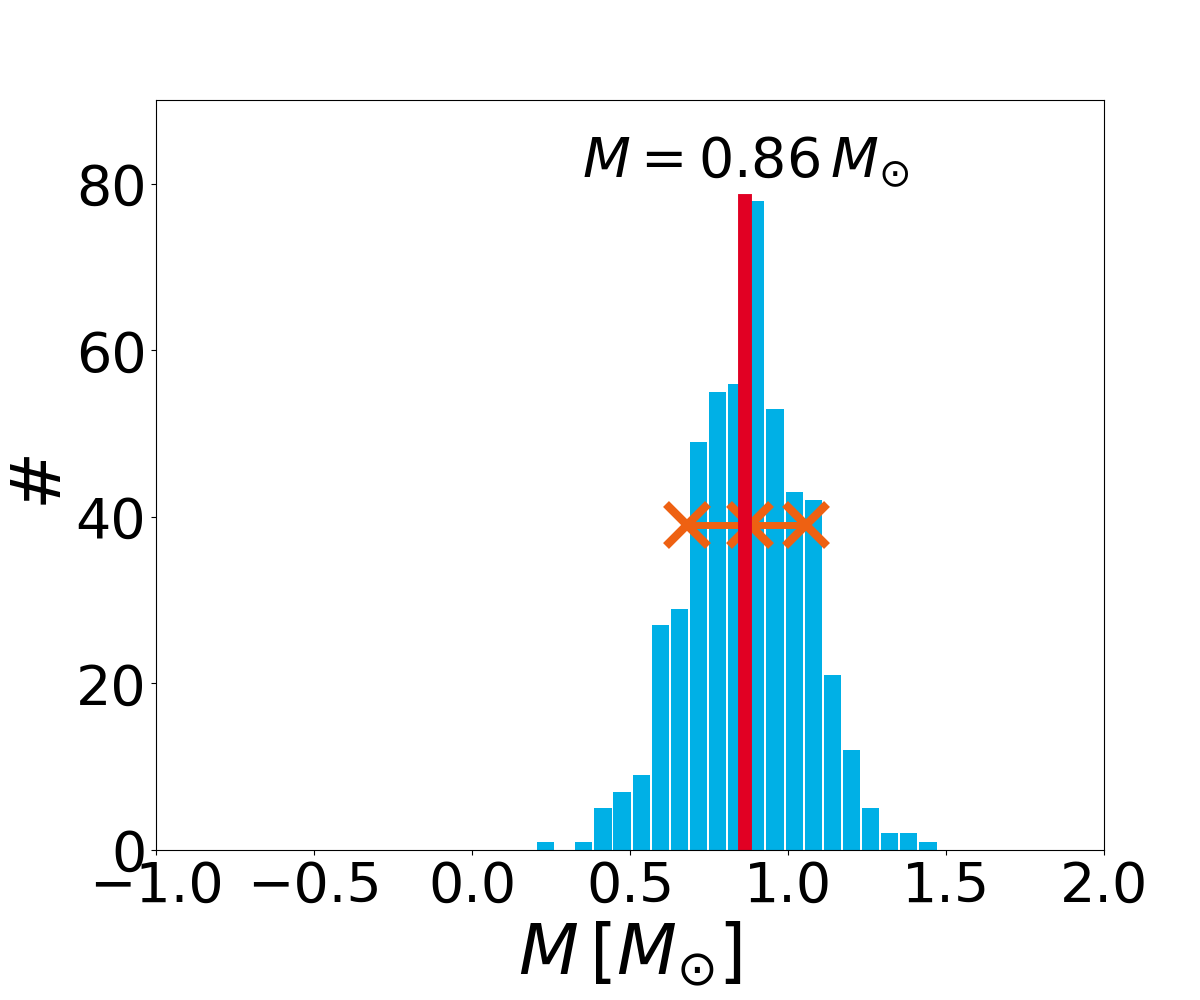}
\caption{}
\end{subfigure}%
\begin{subfigure}{.25\textwidth}
\centering
\includegraphics[width = 4.5cm]{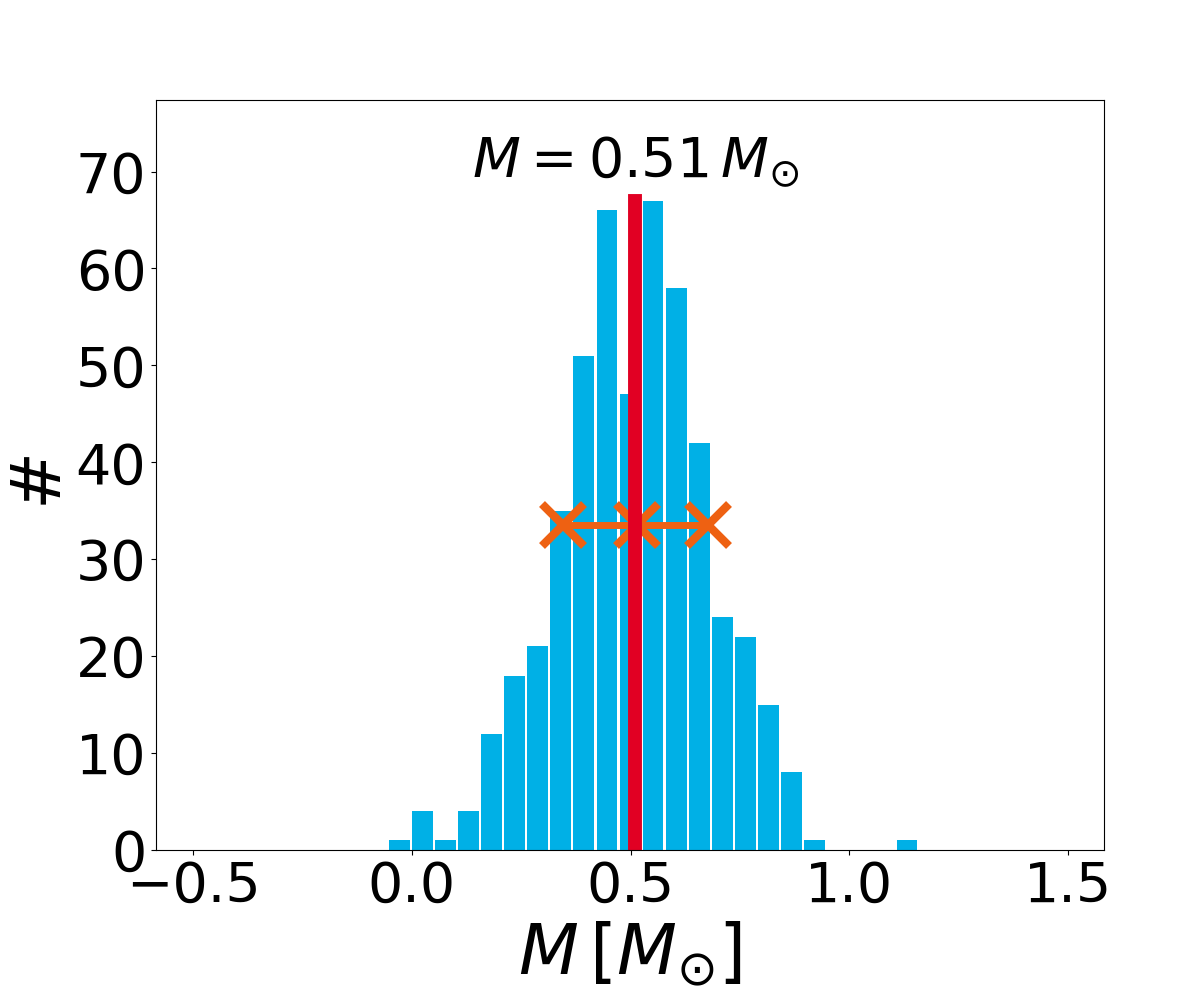}
\caption{}
\end{subfigure}
\begin{subfigure}{.25\textwidth}
\centering
\includegraphics[width = 4.5cm]{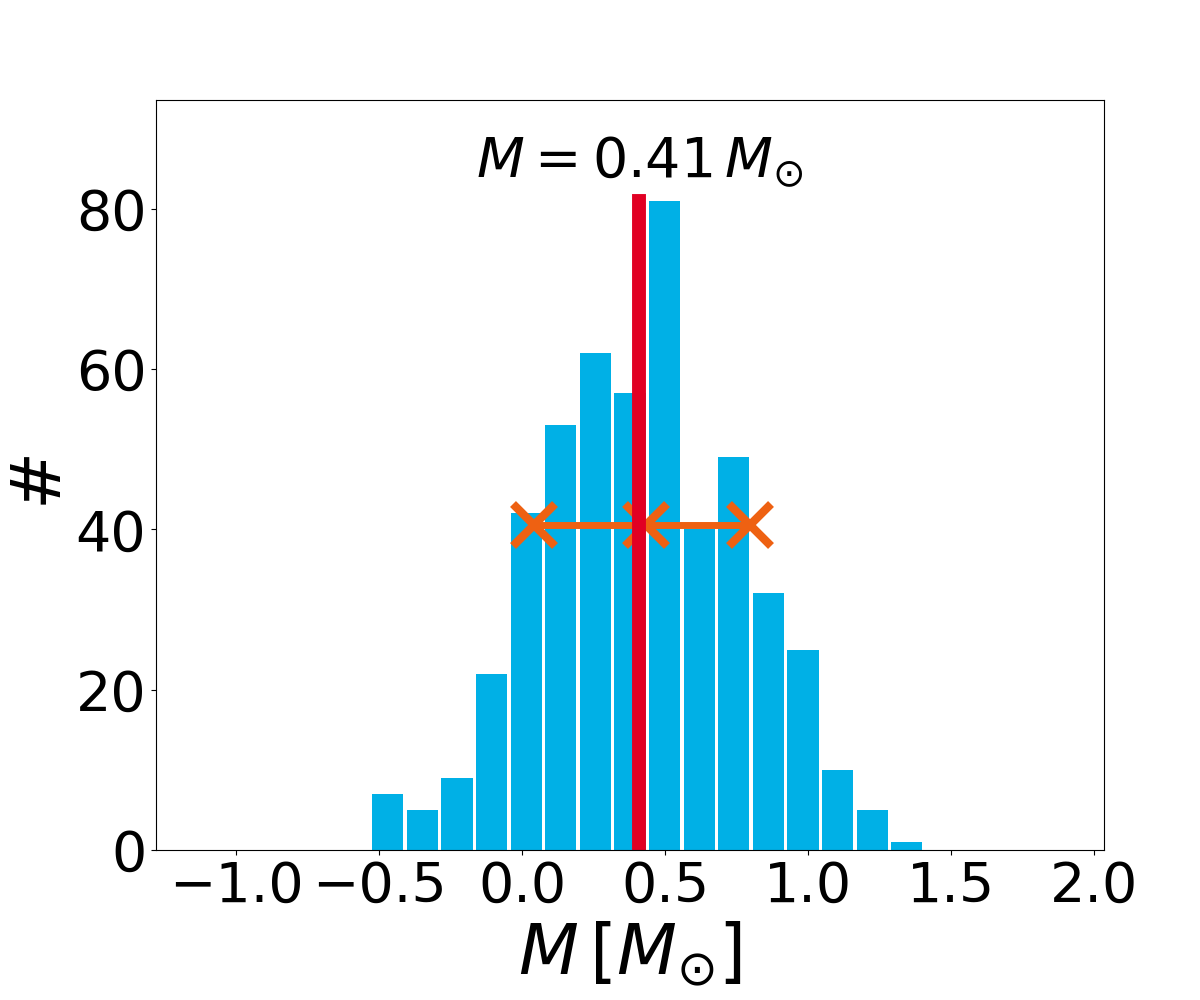}
\caption{}
\end{subfigure}%
\begin{subfigure}{.25\textwidth}
\centering
\includegraphics[width = 4.5cm]{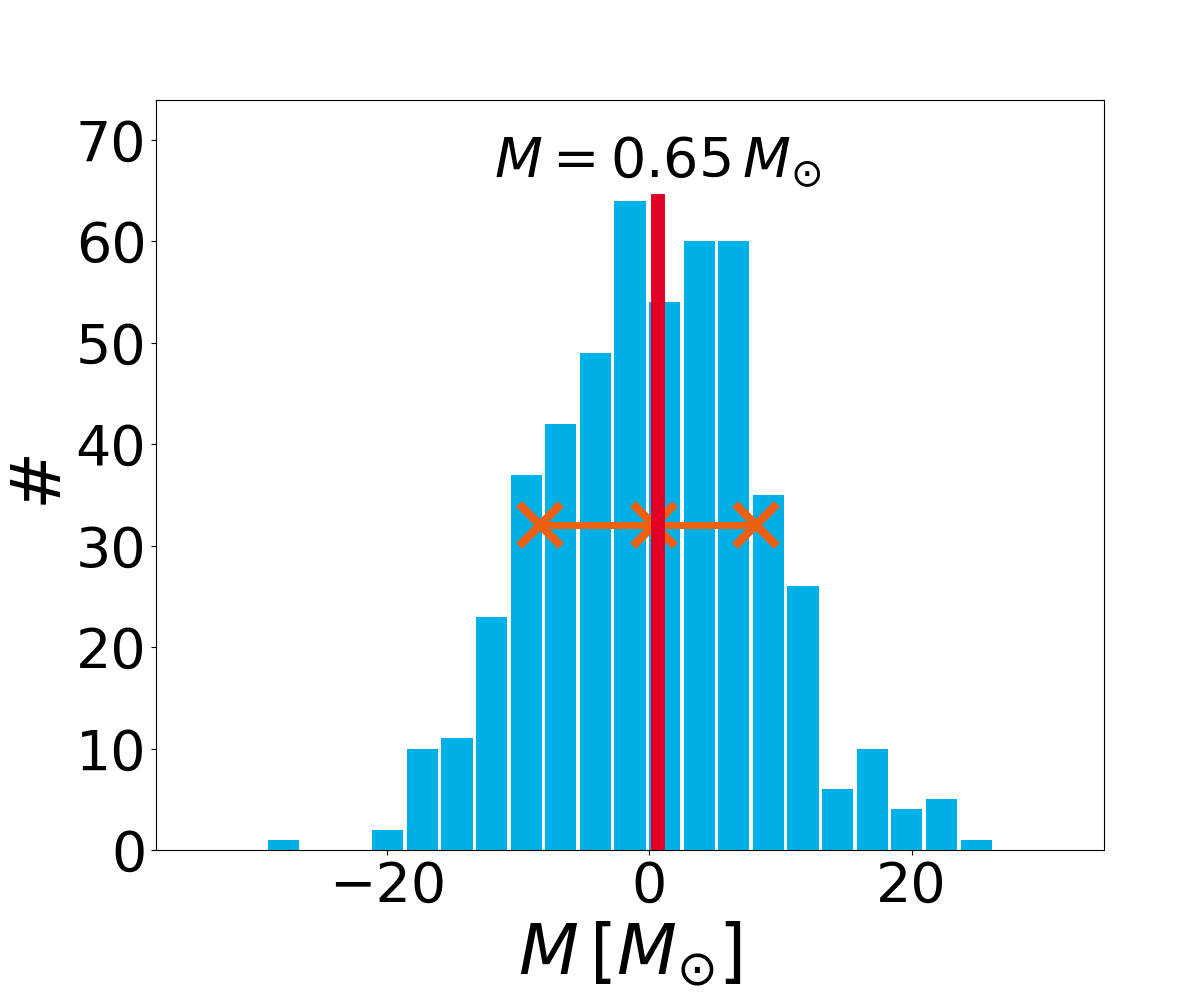}
\caption{}
\end{subfigure}
\caption{Histogram of the simulated mass determination for four different cases:
(a) and (b) precision of about 15\% and 30\%, respectively; \Gaia{} is able to measure the mass of the lens; 
(c) precision between 50\% and 100\%;  
for these events \Gaia{} can detect a deflection, but a good mass determination is not possible;
(d) the scatter is larger than the mass of the lens; \Gaia{} is not able to detect a deflection of the background source.
The orange crosses show the 15.8th, 50th and 84.2th percentiles  (\(1\sigma\) confidence interval) of the 500 realisations, and the red vertical line indicates the input mass. 
Note the much wider \mbox{x-scale} for case (d)!}
\label{figure:mass_distributuions}
\end{figure}

In order to determine the precision of the mass determination we use a Monte Carlo approach. 
We first create a set of error-free data points using the astrometric parameters provided by 
\Gaia{} and the approximated mass of the lens based on the G magnitude estimated by \citet{2018A&A...620A.175K}.  We then create 500 sets of observations, by randomly picking values from the error ellipse of each data point.
We also include a 5\%  chance that a data point is missing, or is clipped as outlier. 
From the sample of 500 reconstructed masses, we determine the 15.8, 50, and 84.2 percentiles  (see Fig.\,\ref{figure:mass_distributuions}). These represent the \(1\sigma\) confidence interval.
We note that a real observation will give us one value from the determined distribution and not necessarily a value close to the true value or close to the median value. 
But the standard deviation of this distribution will be similar to the error of real measurements. 
Further, the median value gives us an insight if we can reconstruct the correct value.  

To determine the influence of the input parameters,  
we repeat this process 100 times while varying the input parameters, that means the positions, proper motions and parallaxes of the lens and source as well as the mass of the lens,  within the individual error distributions. 
This additional analysis is only done for events were the first analysis using the error-free values from \Gaia{} DR2 lead to a \(1\sigma\) uncertainty smaller  than the assumed mass of the lens.

\section{Results} 
 \label{chapter:Result}

\begin{figure}
\includegraphics[width=9.1cm]{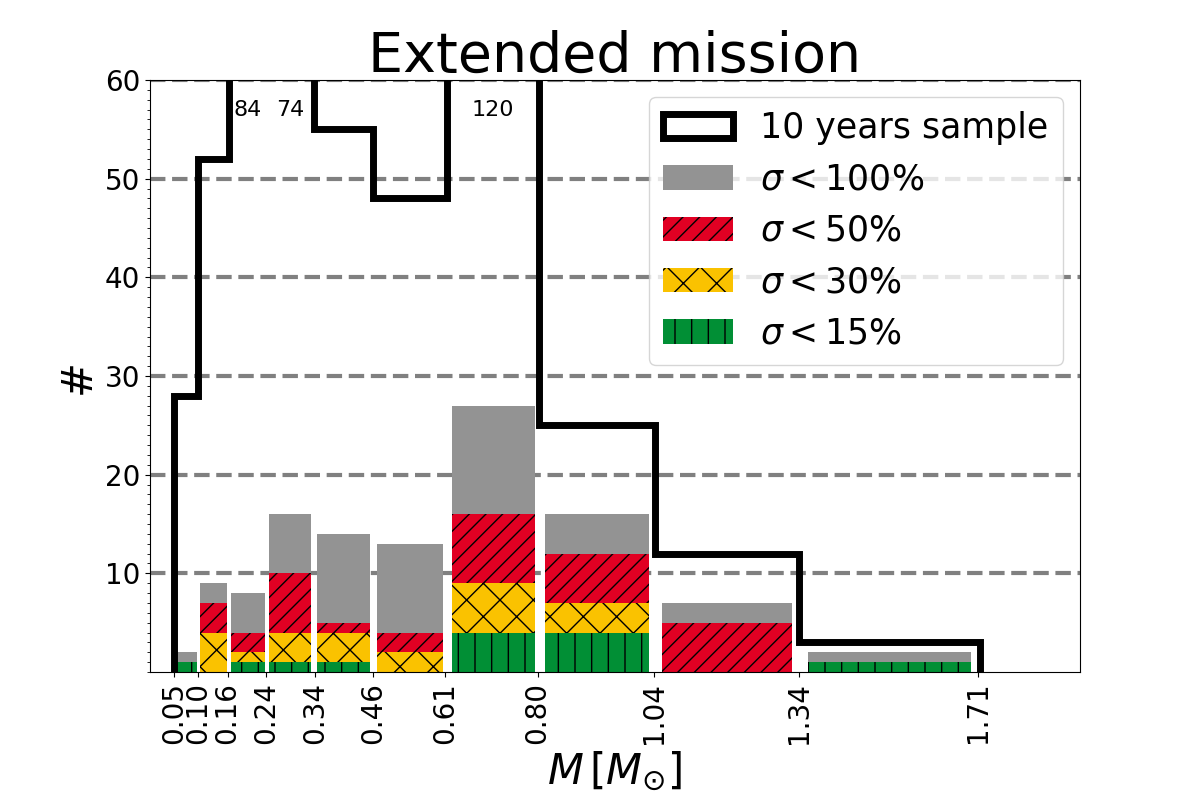}
\includegraphics[width=9.1cm]{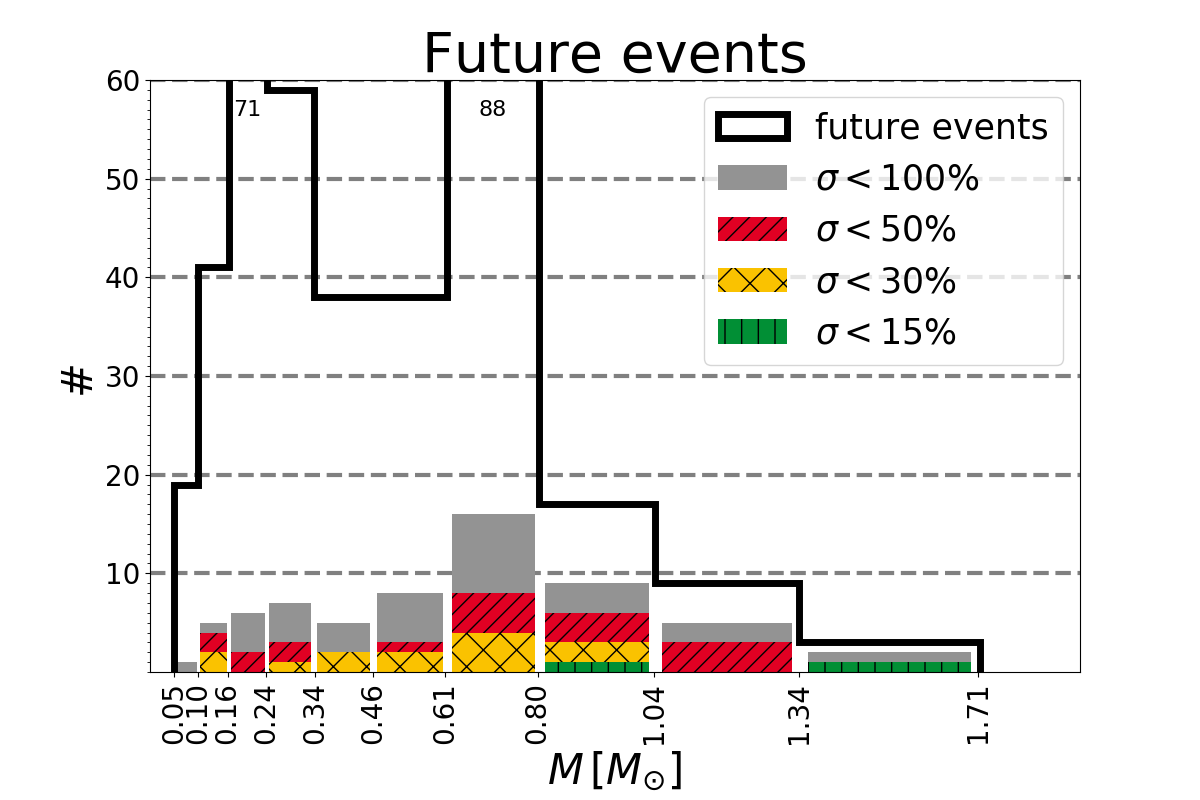}
\includegraphics[width=9.1cm]{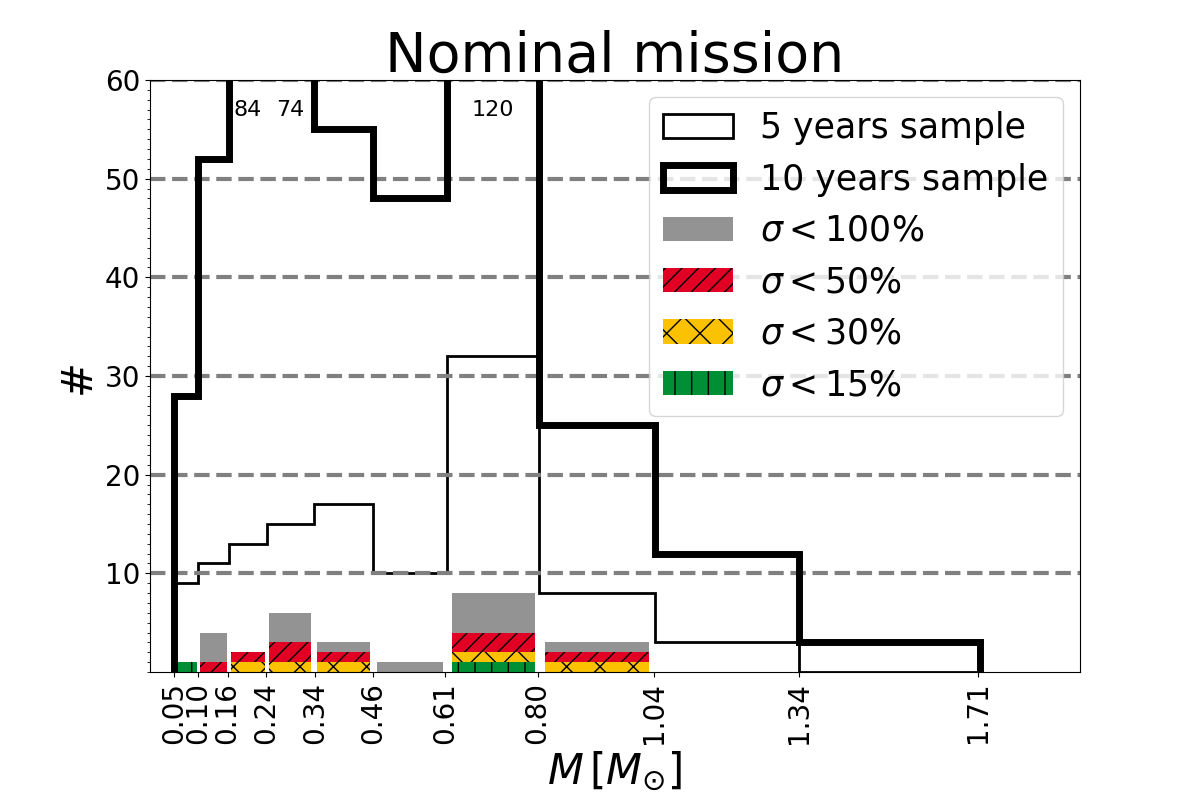}
\caption{Distribution of the assumed masses and the resulting relative standard error of the mass determination  for the investigated events. Top panel: Using the data of the extended 10-year 
mission. 
Middle panel: Events with a closest approach after mid 2019.   
Bottom panel: Using only the data of the nominal 5-year mission.  
The grey, red, yellow  and green parts correspond to a relative standard error better than 100\%, 50\%, 30\% and  15\%. The thick black line shows the distribution 
of the input sample, where the numbers at top show the number of events  in the corresponding bins. The thin black line in the bottom panel shows the events during the nominal mission.
The peak at \(0.65\Msun{}\) is caused by the sample of white dwarfs. The bin size increases by a constant factor of 1.25 from bin to bin.}
\label{figure:hist_mass_accuracy}
\end{figure}
 
\subsection{Single background source}
Using the method described in Section 4, we determine the scatter of individual fits. 
The scatter gives us an insight into the reachable precision of the mass determination using the 
individual \Gaia{} measurements. In our analysis we find three different types of distribution. For 
each of these a representative case is shown in Figure\,\ref{figure:mass_distributuions}. 
For the first two events (Fig.\,\ref{figure:mass_distributuions}\,(a) and \ref{figure:mass_distributuions}\,(b)),  the width of the distributions, calculated via  
50th percentile minus 15.8th percentile and 84.2th percentile minus 50th percentile is smaller than 15\%  and 30\%
of the assumed mass, respectively. For such events it will be possible to determine the mass of the lens once the data are 
released. For the  event of Figure\,\ref{figure:mass_distributuions} (c)  
the standard error is in the same order as the mass itself. 
For such events the \Gaia{} data are affected by astrometric microlensing, however the data are not good enough to determine a precise mass. By including further data, for example observations by the 
Hubble Space Telescope (HST), during the peak of the event, a good mass determination might be possible. This is of special interest for upcoming events in the next years. 
If the scatter is much larger than the mass itself as in Figure\,\ref{figure:mass_distributuions}\,(d)),    
the mass cannot be determined using the \Gaia{} data.

\begin{figure}
\includegraphics[width=9.1cm]{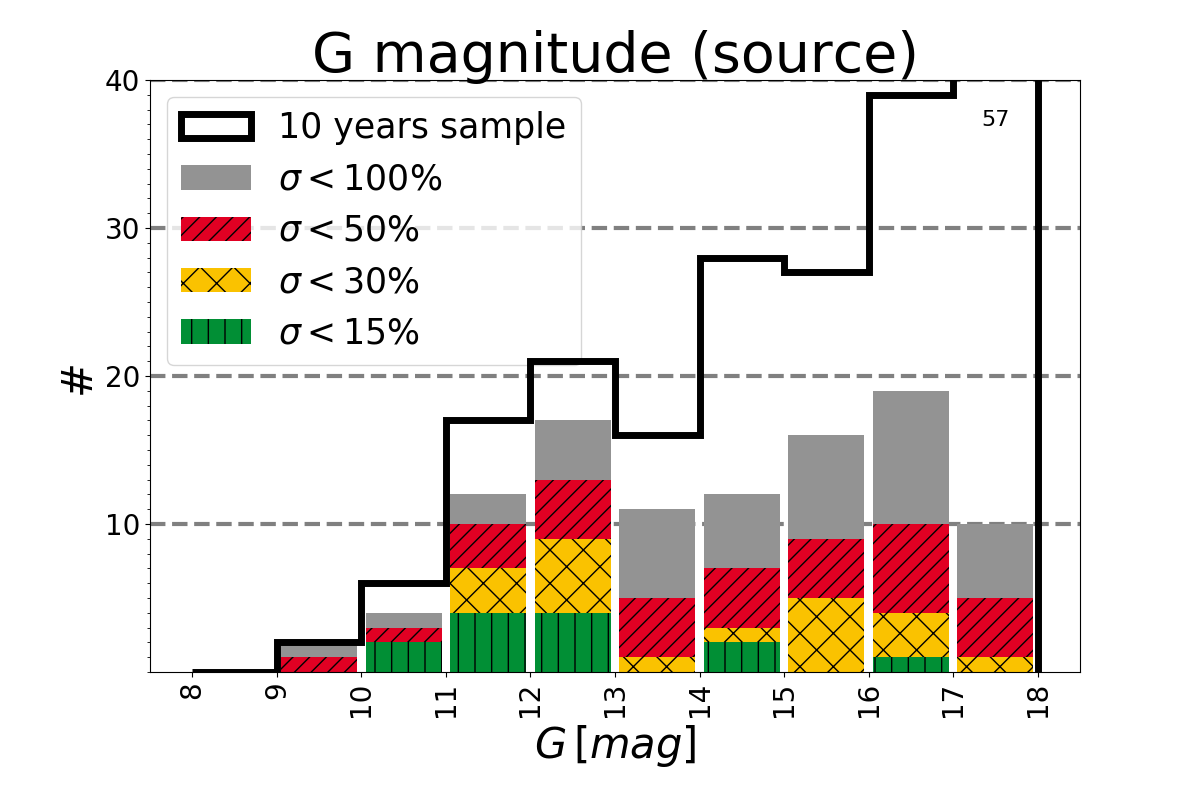}
\includegraphics[width=9.1cm]{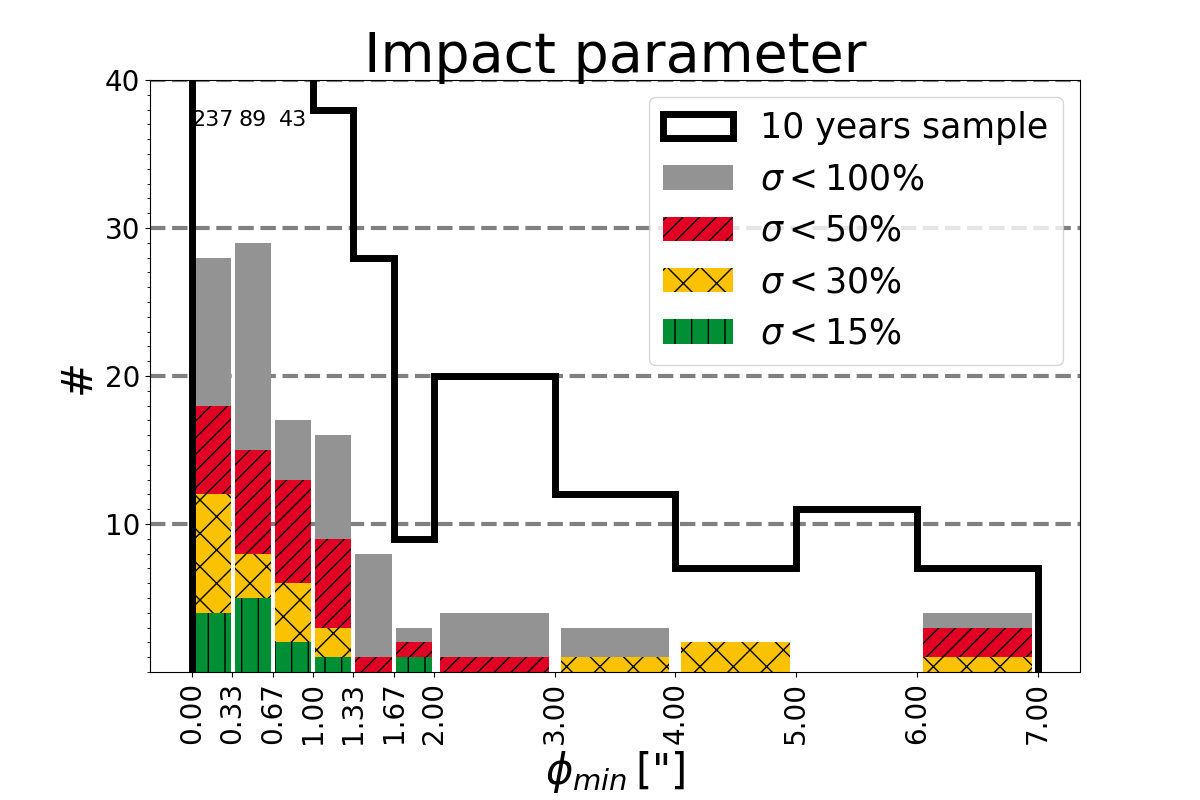}

\caption{Distribution of the G magnitude of the source (top) and  impact parameter (bottom) as well as the resulting relative standard error of the mass determination  for the investigated events.
The grey, red, yellow  and green parts correspond to a relative standard error better than 100\%, 50\%, 30\% and  15\%, respectively. The thick black line shows the distribution 
of the input sample, where the numbers at top show the number of events  in the corresponding bins. The thin black line in the bottom panel shows the events during the nominal mission. Note the different bin width of \(0.33''\) below \(\phi_{min} = 2''\) and \(1''\) above \(\phi_{min} =  2''\) in the bottom panel.}
\label{figure:hist_GS_accuracy}
\end{figure}

In this analysis, we test 501 microlensing events, predicted for the epoch J2014.5 until J2026.5 by  \cite{2018A&A...620A.175K}. 
Using data for the potential 10-year extended \Gaia{} mission, we find that the mass of 13  lenses 
can be reconstructed with a relative uncertainty of \(15\%\) or better.  Further 21 events can be reconstructed with a 
relative standard uncertainty better than \(30\%\) and additional 31 events with an uncertainty better than \(50\%\) ( i.e. 
\(13+21+31 = 65\) events can be reconstructed with an uncertainty smaller than  \(50\%\) of the mass). 
The percentage of events where we can reconstruct the mass increases with the mass of the lens
(see Fig.\,\ref{figure:hist_mass_accuracy}). This is not surprising since a larger lens mass results in a larger microlensing effect. Nevertheless, with \Gaia{} data it is also possible to derive the masses  of the some low-mass stars \((M< 0.65\Msun{})\)  with a small relative error (\(<15\%\)). It is also not surprising that for brighter background sources it is easier to reconstruct the mass of the event (Fig.~\ref{figure:hist_GS_accuracy} top panel), due to the better precision of \Gaia{} (see Fig.~\ref{figure:sig_gmag}). Further, the impact parameters of the reconstructable events are typically below \(1.67"\) with a peak around \(0.35''\). This is caused by the size of \Gaia{}'s readout windows.
Using only the data of the nominal 5-year mission  we can observe the same trend. However, due 
to the fewer data points and the fact that most of the events reach the maximal deflection after the end of the nominal mission (2019.5), the fraction of events with a certain relative uncertainty of the 
mass reconstruction is much smaller (see Fig.\,\ref{figure:hist_mass_accuracy} bottom panel). So the 
mass can only be determined for \(1\), \(1+ 6 = 7\),  and \(1+ 6+8 = 15\) events  with an 
uncertainty better than \(15\%\), \(30\%\) and \(50\%\), respectively.

\begin{figure}
\includegraphics[width=9cm]{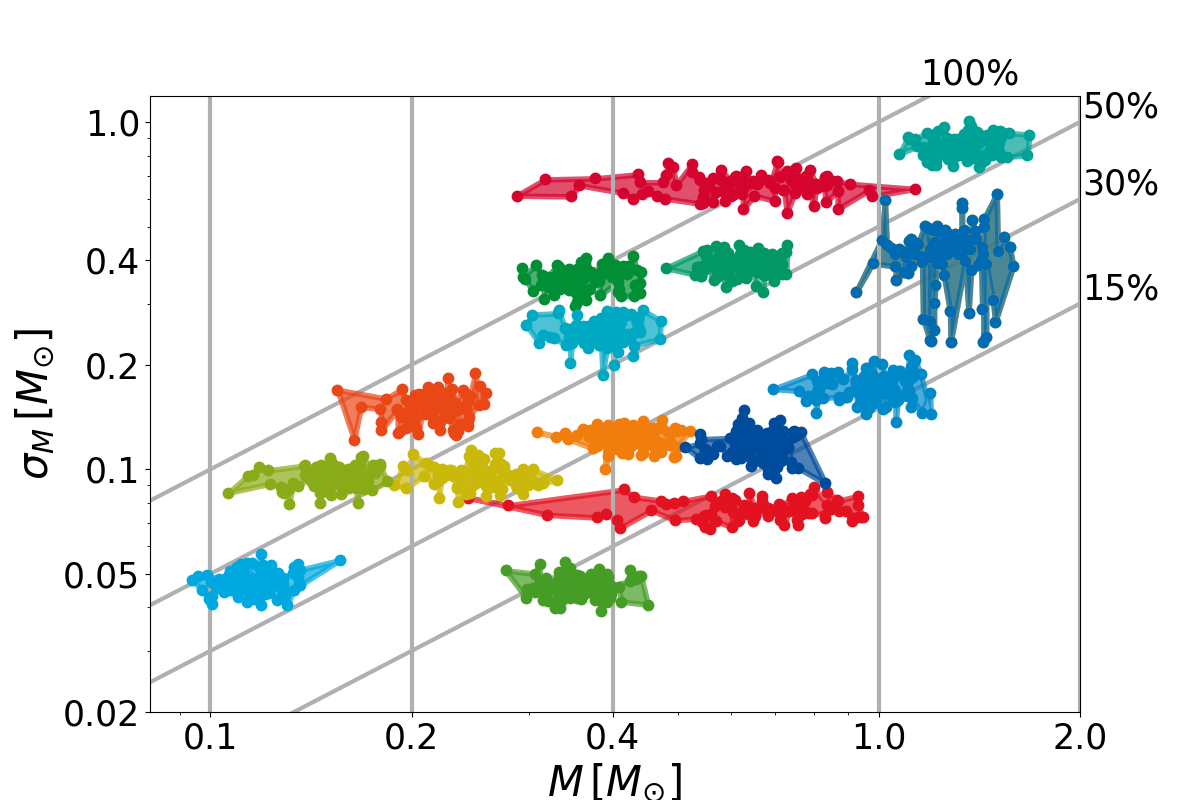}
\caption{\label{figure:plot_mass_accuracy}
Achievable standard error as a function of the input mass for 15 events. The two red events with a wide range for the input mass are white dwarfs, were the mass can only poorly be determined from the G magnitude. The uncertainty is roughly constant as function of the input mass.
The diagonal lines indicates relative uncertainty of 15\%, 30\%, 50\% and 100\%, respectively.}
\end{figure}

For 114  events, where the expected uncertainty is smaller than \(100\%\), we expect that \Gaia{} can at least qualitatively detect the astrometric deflection. For those we  
repeat the analysis while varying the input parameters for the data simulation. 
Figure\,\ref{figure:plot_mass_accuracy}
shows the achievable precision as a function of the input mass for a representative subsample. When the proper motion of the background star is known from \Gaia{} DR2, the uncertainty of the achievable precision is about \(6\%\), and it is about \(10\%\) when the proper motion is unknown. 
We find that the reachable uncertainty (in solar masses) depends only weakly on the input mass, and 
is more connected to the impact parameter, which is a function of all astrometric input parameters. Hence, the scatter of the achievable 
precision is smaller when the  proper motion and parallax of the background source is known from
\Gaia{} DR2. 
For the 65 events with a relative standard error better than 50\%, Table\,\ref{table:single} and Table\,\ref{table:single2}  list the achievable relative uncertainty for each individual star as well as the determined scatter, for the  extended mission \(\sigma M_{10}\). 
Table\,\ref{table:single} contains all events during the nominal mission (before 2019.5), and also includes the  determined scatter using only the data of the nominal mission \(\sigma M_{5}\). Table\,\ref{table:single2} lists all  future events with a closest approach after 2019.5.

\subsection*{Future events}
In our sample, {383} events have a closest approach after 2019.5 
(Fig.\,\ref{figure:hist_mass_accuracy} middle panel).
These events are of special interest, since it is possible to obtain further observations using 
other telescopes, and to combine the data. Naively, one might expect that about 50\% of the events should be after this 
date (assuming a constant event-rate per year). However, the events with a closest approach close to the 
epochs used for \Gaia{} DR2 are more difficult to treat by the \Gaia{} reduction  (e.g. fewer observations 
due to blending). Therefore many background sources are not included in \Gaia{} DR2. 
For 17 of these future events, the achievable relative uncertainty is between 30 and 50 per cent. Hence the  
combination with further precise measurements around the closest approach is needed to determine a precise mass of the lens.
To investigate the possible benefits of additional observations, we repeated the simulation while adding two  2-dimensional observations (each consists of two perpendicular 1D observations) around the epoch of the closest approach. We only consider epochs, where the separation between the source and lens stars is larger than \(150\,\mathrm{mas}\).
Further, we assume that these observations have the same precision as the \Gaia{} observations. These results are listed in the column \(\sigma M_{obs}/M_{in}\) of Table\,\ref{table:single2}.
\\
By including these external observation the results can be improved by typically 2 to 5 percentage points. In extreme cases, the value can even  even a factor 2 when the impact parameter is below \(0.5''\), since \Gaia{} will lose measurements due to combined readout windows.
The events \(\#63\) to \(\#65\) are special cases, since they are outside the extended mission, and \Gaia{} only observes the leading tail of the event.

\subsection{Multiple background sources} 
\begin{figure}
\centering
\begin{subfigure}{.5\textwidth}
\center
\includegraphics[width =8.35cm]{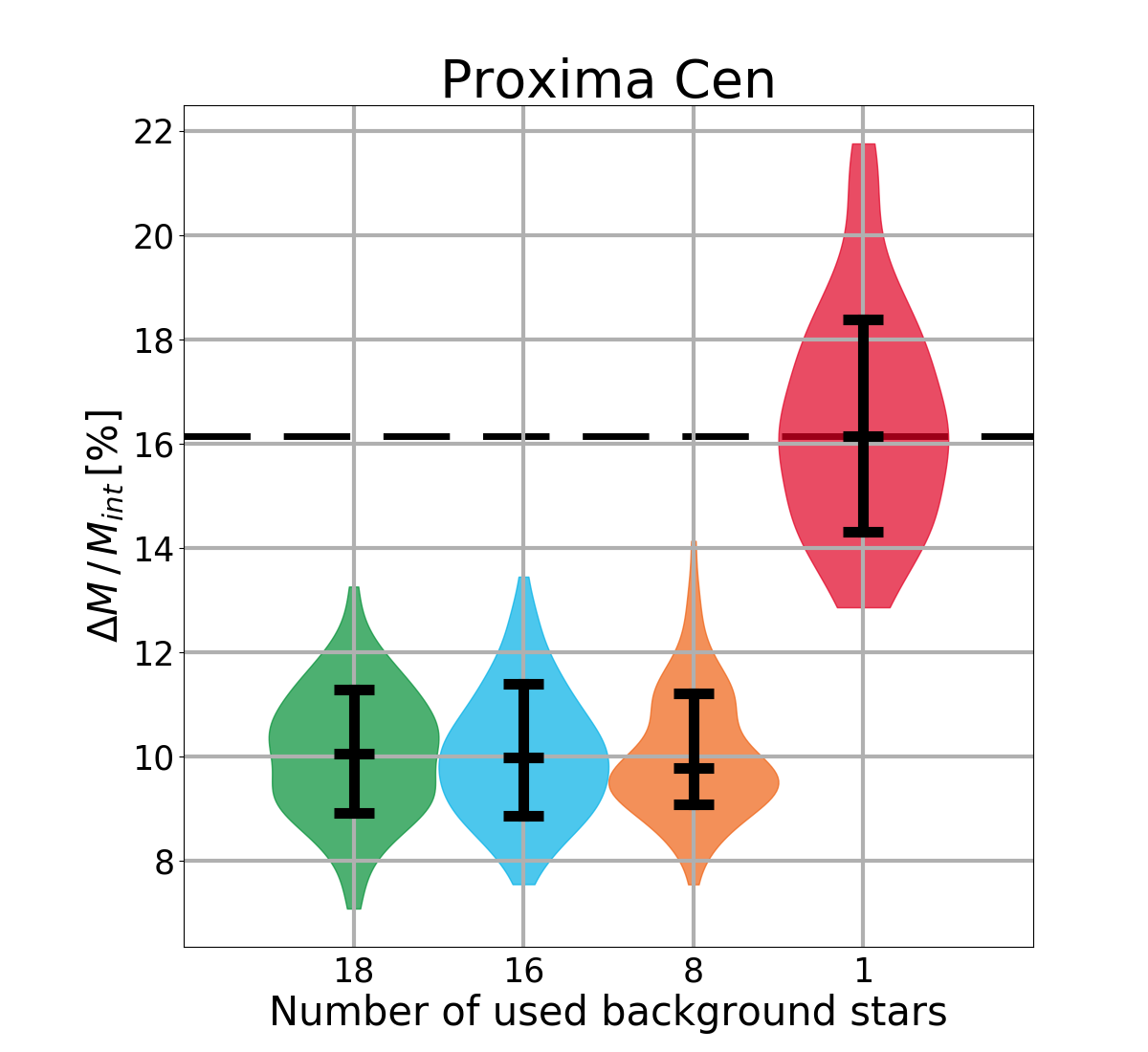}
\end{subfigure}
\begin{subfigure}{.5\textwidth}
\center
\includegraphics[width =8.35cm]{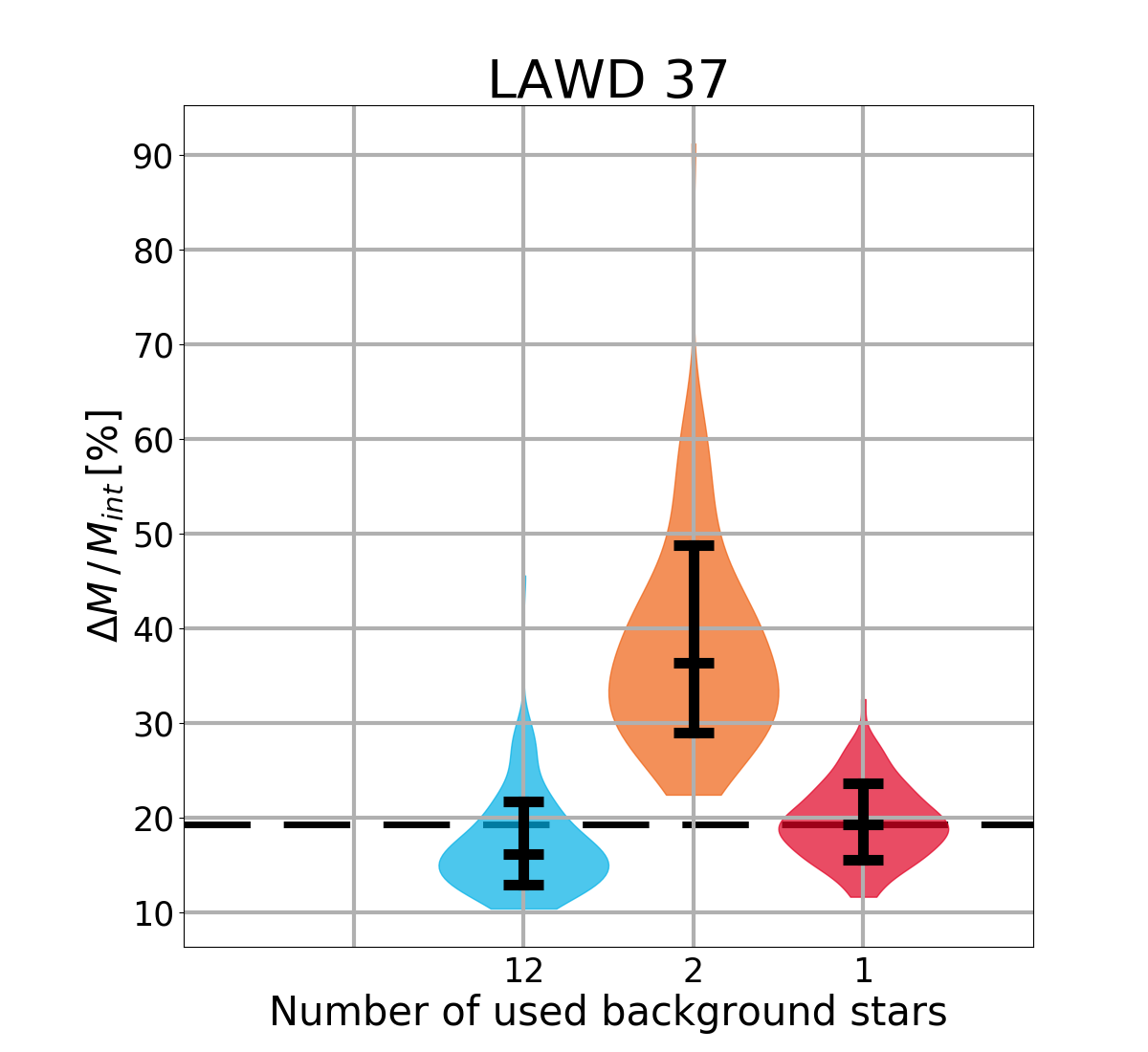}
\end{subfigure}%
\caption{
Violin plot\textcolor{blue}{\protect\footnotemark[10]} of the achievable uncertainties for the four different methods for Proxima Centauri (top) and  LAWD 37 (bottom). 
For each method the 16th, 50th and 84th, percentile are shown. The shape shows the distribution of the 100 determined uncertainties from vary the input parameter. This distribution is smoothed with a Gaussian kernel.  
The green ``violins'' use all of the background sources. 
For the blue ``violins''  only background sources with a 5-parameter solution are used, 
and for the orange ``violins'' only stars with a  precision in along-scan direction better than \(0.5\,\mathrm{mas}\) and a 5-parameter solution are used. 
The red ``violins''  shows the best results when only one source is used. The dashed line indicates the median of this distribution.
For each method the number of used stars is list below the ``violin''. 
The missing green ``violin''  of LAWD 37 is caused by no additional background stars with a 2-parameter solution only. Hence it would be identical to the blue one
(For the other events with multiple background stars see Fig.\,\ref{figure:diff_ideas_all} in the appendix)}
\label{figure:diff_ideas_prox} 
\end{figure}

\begin{figure*}
\begin{subfigure}{.33\textwidth}
\centering
\includegraphics[width = 6cm]{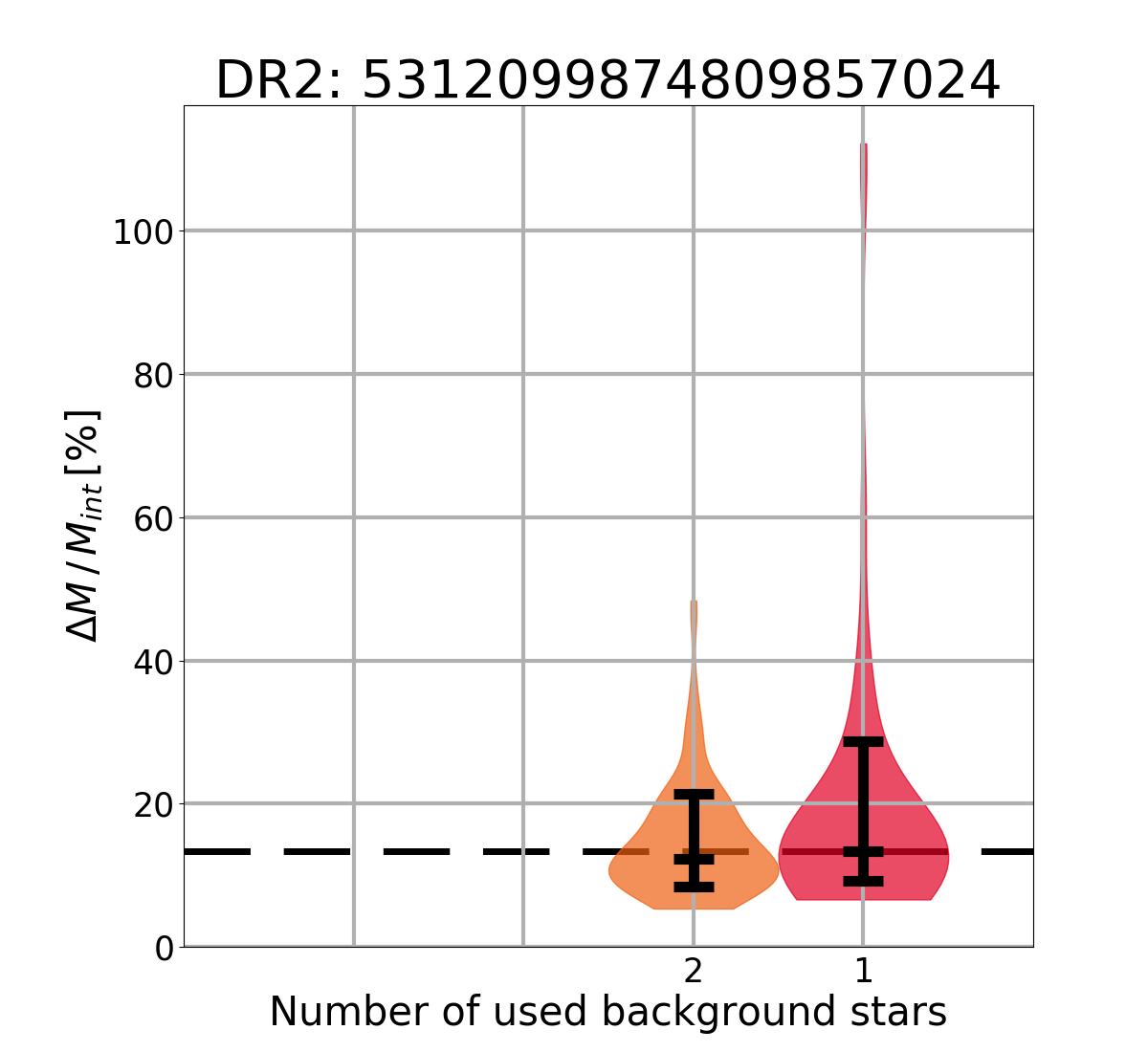}
\caption{}
\end{subfigure}%
\begin{subfigure}{.33\textwidth}
\centering
\includegraphics[width = 6cm]{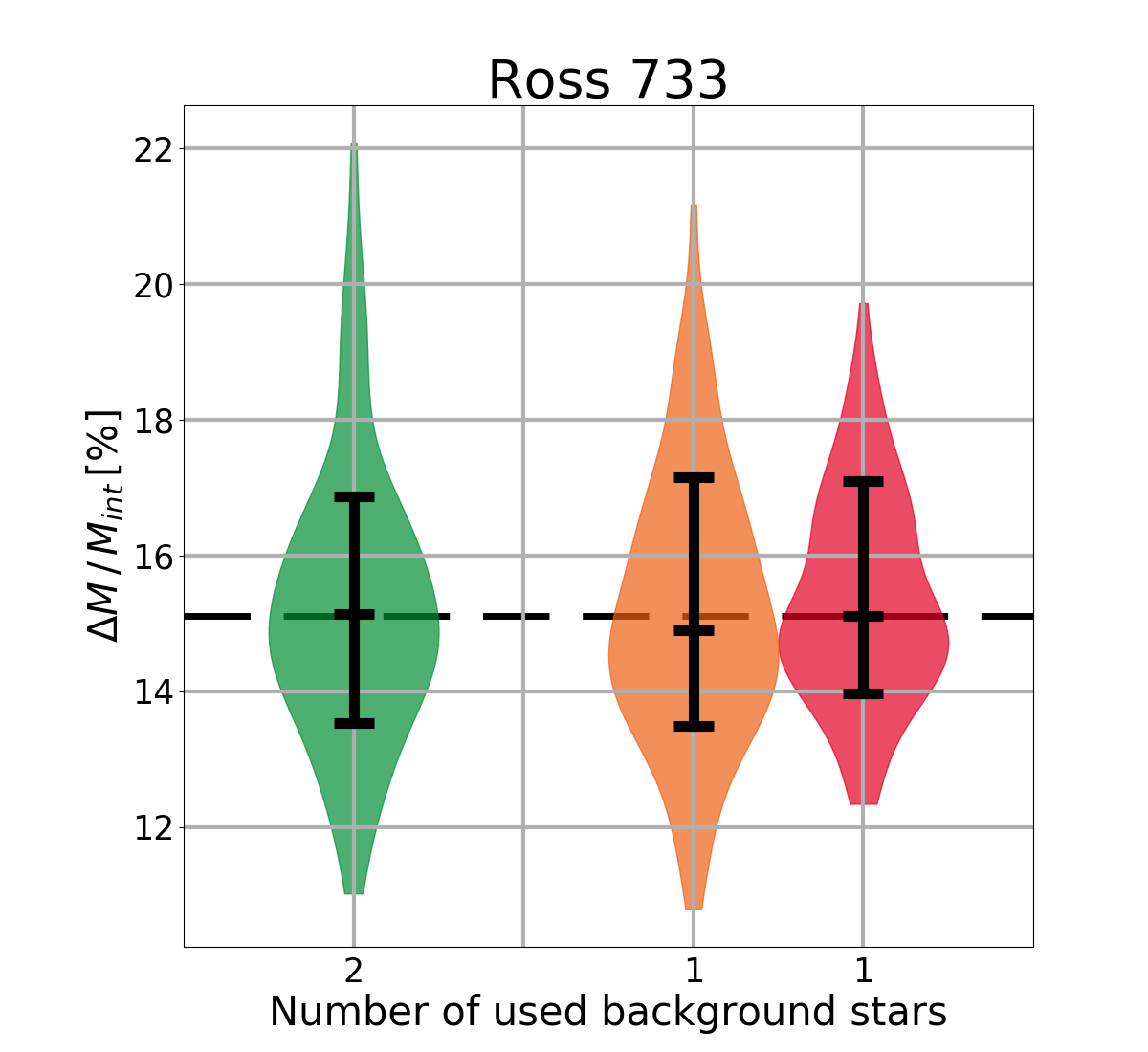}
\caption{}
\end{subfigure}%
\begin{subfigure}{.33\textwidth}
\centering
\includegraphics[width = 6cm]{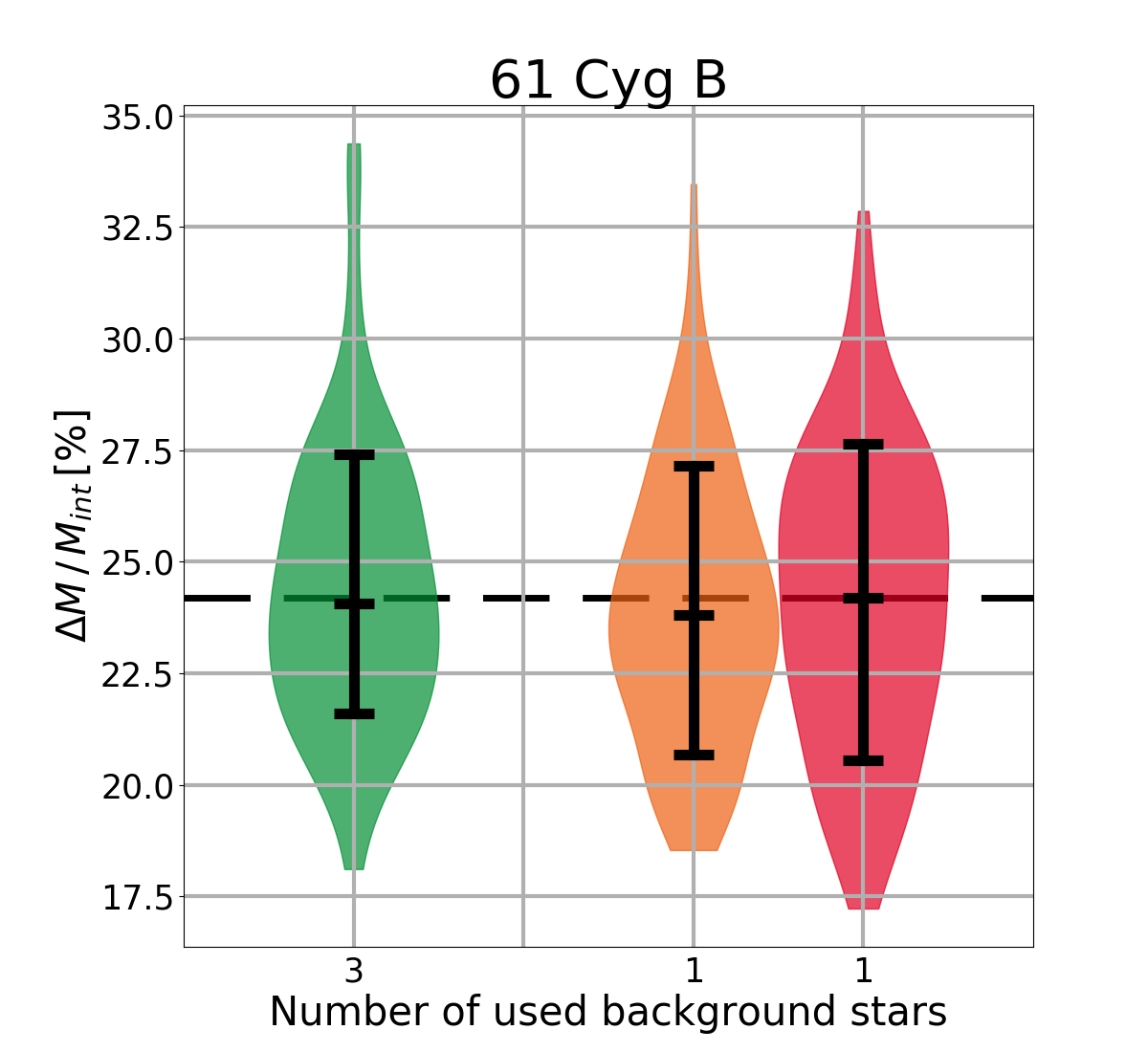}
\caption{}
\end{subfigure}
\begin{subfigure}{.33\textwidth}
\centering
\includegraphics[width = 6cm]{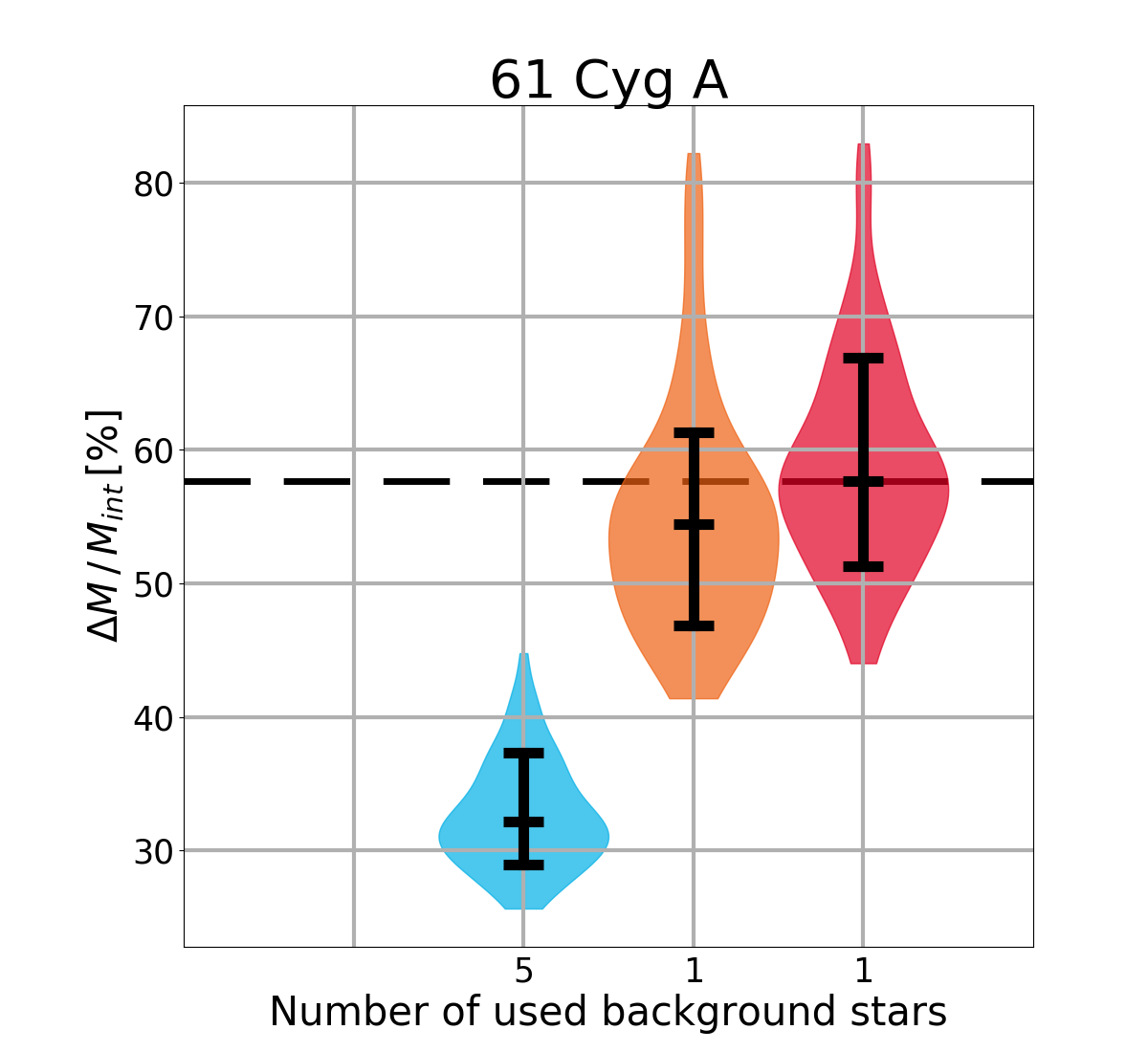}
\caption{}
\end{subfigure}%
\begin{subfigure}{.33\textwidth}
\centering
\includegraphics[width = 6cm]{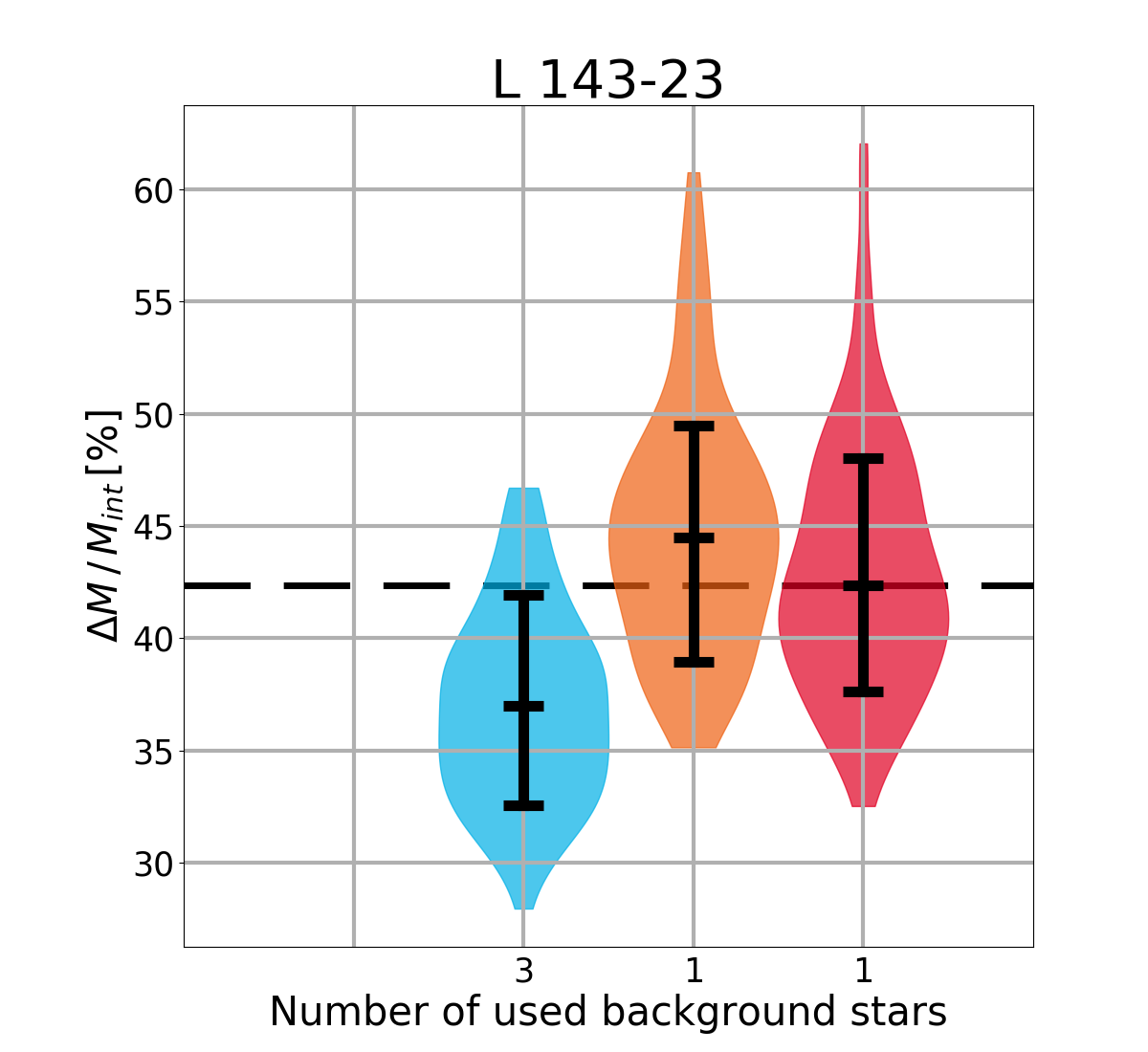}
\caption{}
\end{subfigure}%
\begin{subfigure}{.33\textwidth}
\centering
\includegraphics[width = 6cm]{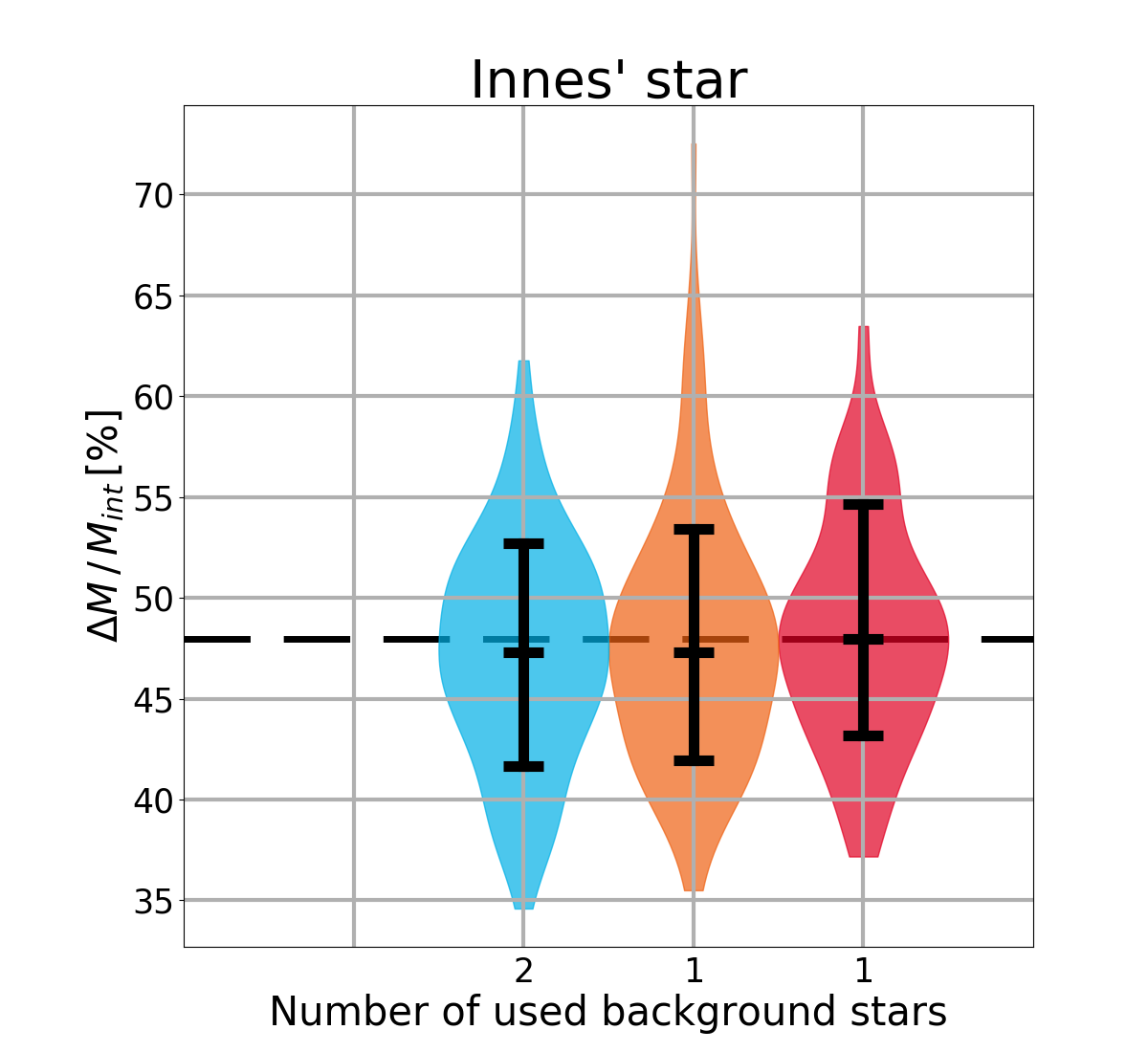}
\caption{}
\end{subfigure}
\begin{subfigure}{.33\textwidth}
\centering
\includegraphics[width = 6cm]{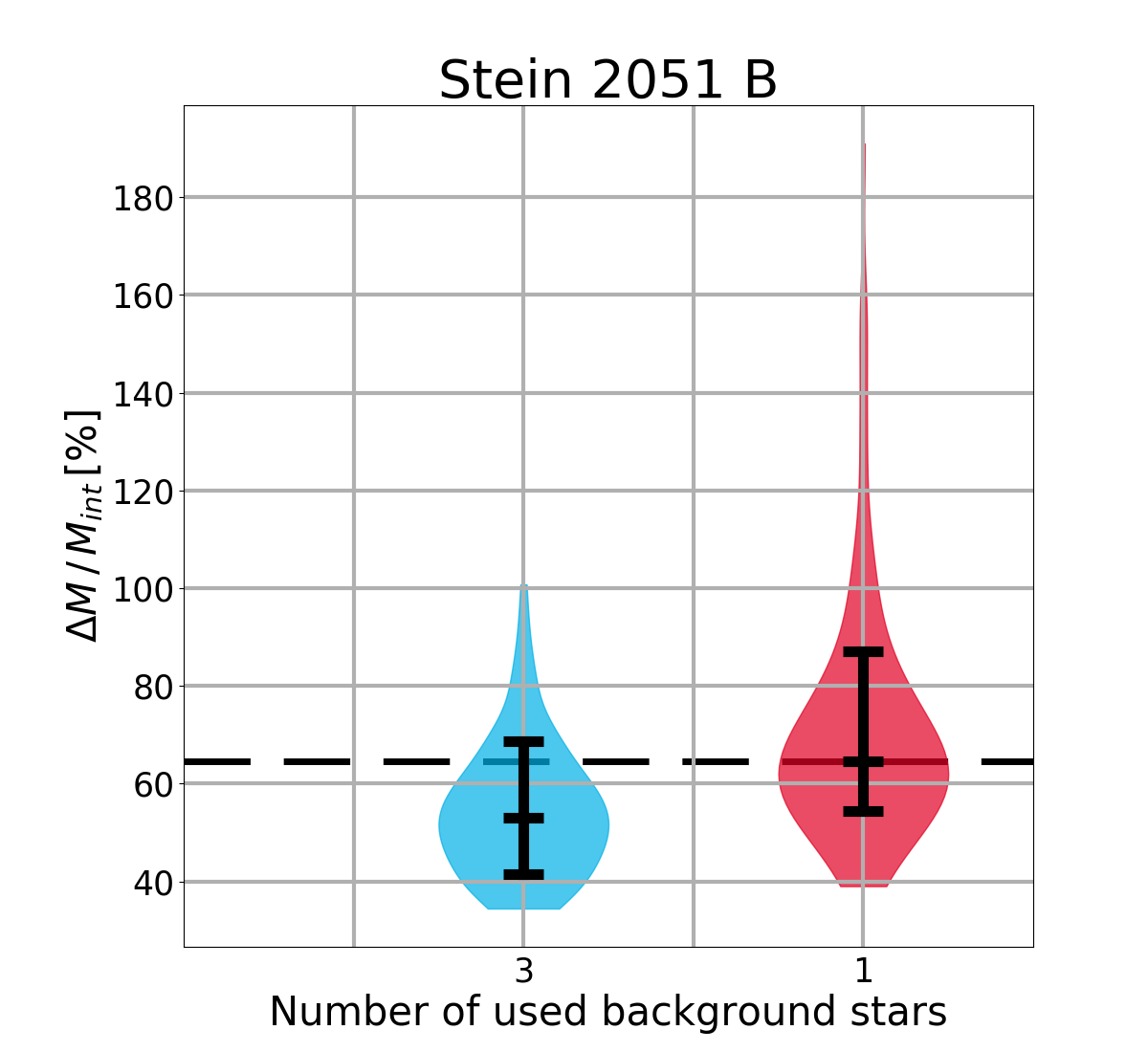}
\caption{}
\end{subfigure}%
\begin{subfigure}{.33\textwidth}
\centering
\includegraphics[width = 6cm]{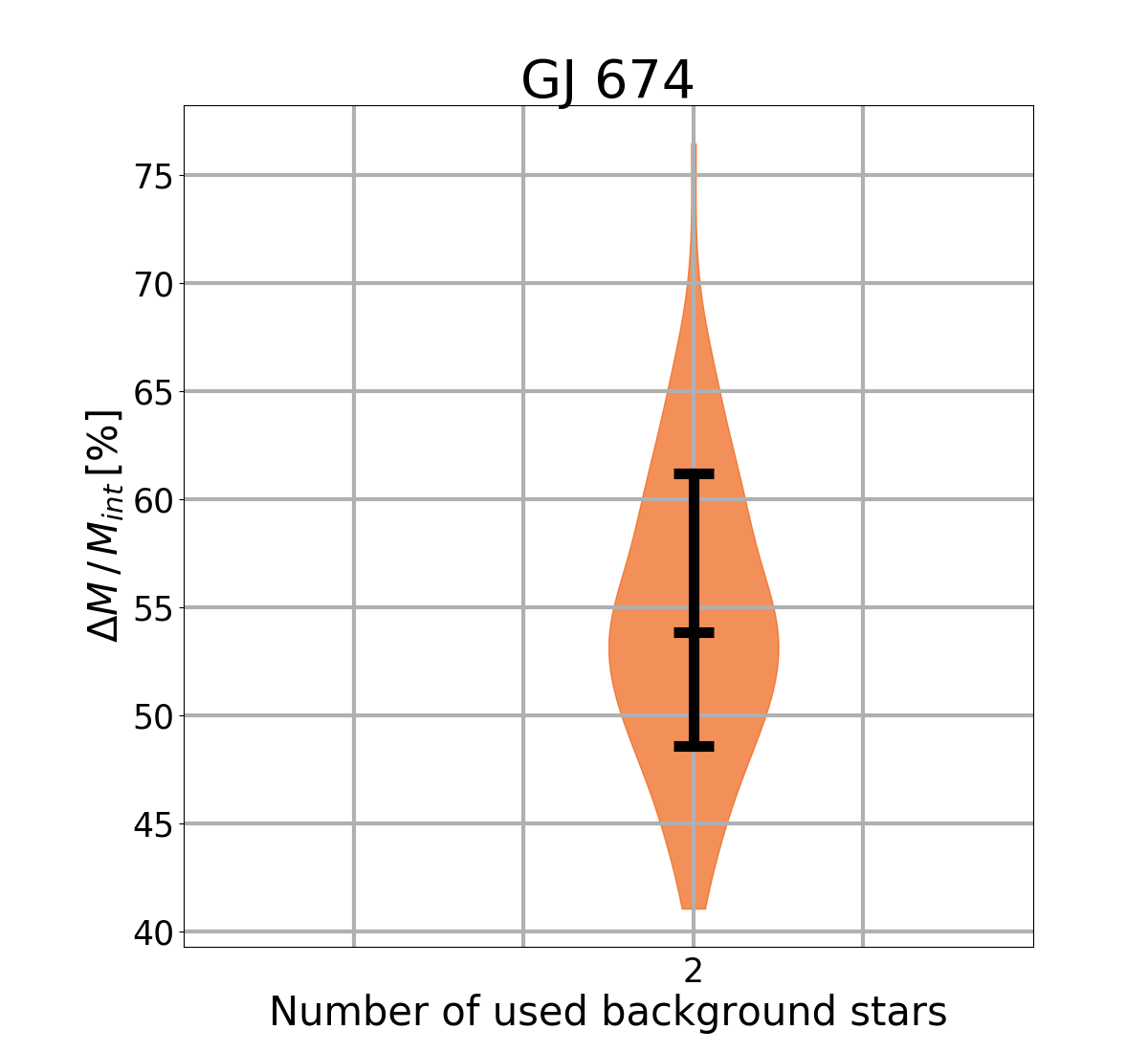}
\caption{}
\end{subfigure}%
\begin{subfigure}{.33\textwidth}
\centering
\includegraphics[width = 6cm]{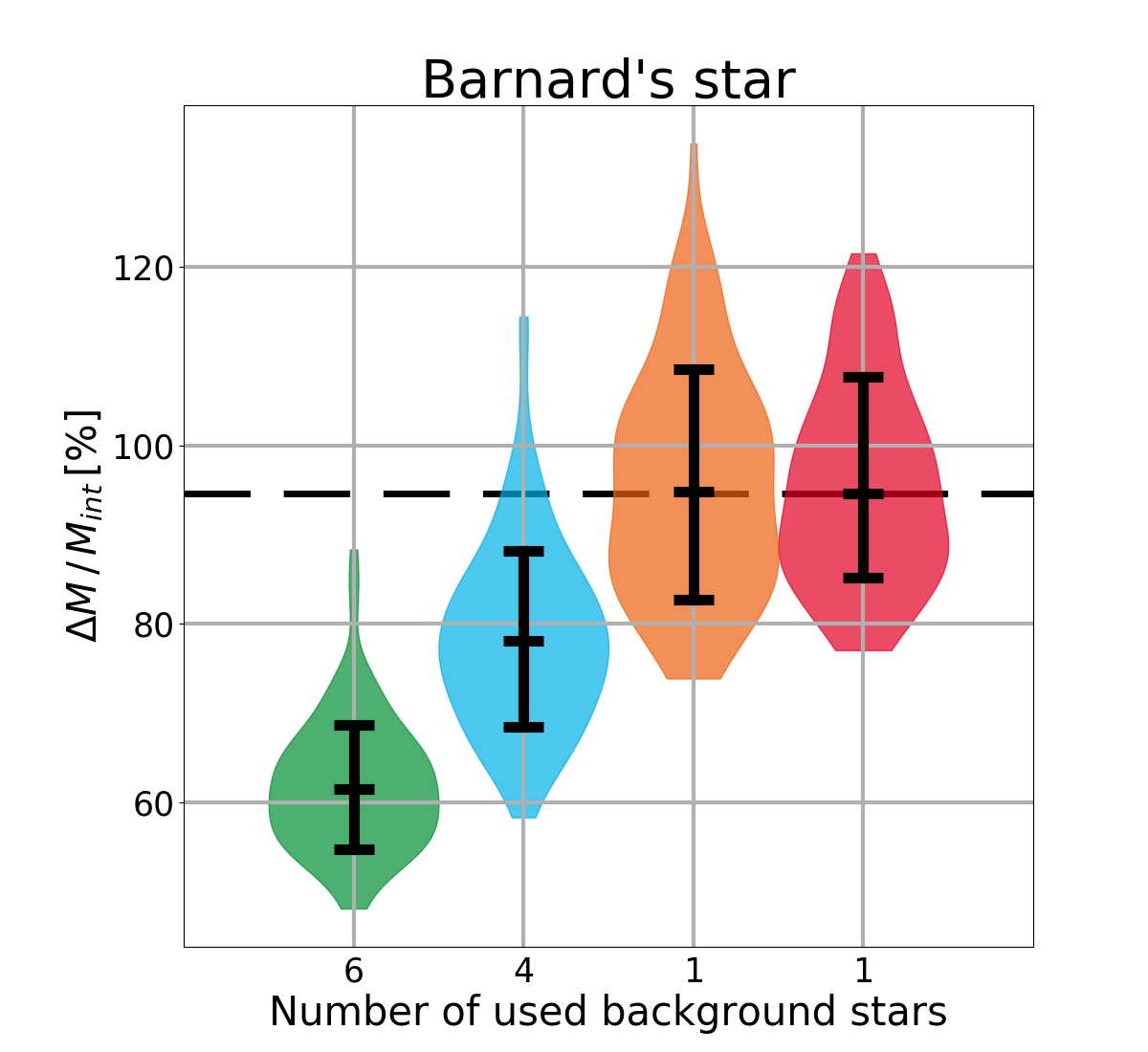}
\caption{}
\end{subfigure}
\caption{
Violin plot of the achievable precision for the four different methods for:
(a) \Gaia{} DR2: 5312099874809857024, (b)\,Ross~733,
(c)\,61~Cyg~B, (d)\,61~Cyg~A, (e)\,L~143-23, (f)\,Innes'~star, (g)\,Stein~2051~B, (h)\,GJ~674 and (i)\,Barnard's~star. 
For each method the 16th, 50th and 84th percentiles are shown. The shape shows the distribution of the 100 determined precisions  smoothed with a Gaussian kernel.  
In each plot, the green ``violin'' uses the all background sources. 
For the blue ``violin'' only background sources with a 5-parameter solution are used, 
and for the orange ``violin'' only stars with a  precision in along-scan direction better than \(0.5\,\mathrm{mas}\)  and a 5-parameter solution are used. 
The red ``violin''  indicates the best results when only one source is used. The dashed line indicates the median of this distribution.
For each method the number of used stars is list below the ``violin''. 
Missing green ``violins''  (e.g. L~143-23\,(e)) are  caused by no additional background stars with a 2-parameter solution only.
Missing blue ``violins'' (e.g. Ross~733\,(b)) are due to the fact that all background sources with a 5-parameter solution have an expected precision in along-scan direction better than \(0.5\,\mathrm{mas}\). For Stein~2051~B\,(g) none of the background stars have an expected precision better than  \(\sigma_{AL} = 0.5\,\mathrm{mas}\), hence the orange ``violin'' is missing. Finally the first analysis of GJ~674\,(h) and Barnard's~star\,(i) using only one background sources results in a precision worse than \(100\%\), consequently the red  ``violins'' are missing. 
}

\label{figure:diff_ideas_all}
\end{figure*}

For the 22 events of \cite{2018A&A...620A.175K} with multiple background sources, we test three different cases: 
Firstly, we use all potential background sources. Secondly, we only use background sources where \Gaia{} 
DR2 provides all 5 astrometric parameters,   
and finally, we select only those background sources for which the expected precision of \Gaia{} individual measurements is better than \(0.5\,\mathrm{mas}\).
The expected relative uncertainties of the mass determinations for the different cases are shown in  Figures\,
\ref{figure:diff_ideas_prox} and \ref{figure:diff_ideas_all}, as well as the expected relative uncertainties for 
the best case using only one background source. 
By using multiple background sources, a better precision of the mass determination can be reached. 
We note that averaging the results of the individual fitted masses will not necessarily increase the precision, since the values are highly correlated.

Using all sources  it is possible to determine the mass of Proxima Centauri with a standard error of \(\sigma_{M} =0.012\Msun{}\) for the extended 10-year mission of \Gaia{}. This corresponds to a relative error of \(10\%\), considering the assumed mass of  \(M = 0.117\Msun{}\)
This is roughly a factor \(\sim0.7\) better than the uncertainty of the best event only (see Fig.\,\ref{figure:diff_ideas_prox} top panel,  \(\sigma_{M} = 0.019\Msun{}\widehat{=}16\%\)).
Since we do not include the potential data points of the two events predicted by \citet{2014ApJ...782...89S}, it might be possible to reach an even higher precision. 
For those two events, \citet{2018MNRAS.480..236Z} measured the deflection using the Very Large Telescope (VLT) equipped with the SPHERE{\protect\footnotemark[9]} instrument. They derived a mass of  \(M=0.150^{+0.062}_{-0.051}\Msun{}\).
Comparing our expectations with their mass determination, we expect to reach a six times smaller error. 

A further source which passes multiple background sources is the white dwarf LAWD 37, where we assume a mass of \(0.65\Msun{}\). Its most promising event, which was first predicted by \cite{2018MNRAS.478L..29M} is in November 2019. \cite{2018MNRAS.478L..29M} also mentioned that \Gaia{} might be able to determine the mass with an accuracy of \(3\%\), However this was done without knowing the scanning law for the extended mission. We expect an uncertainty for the mass determination by \Gaia{} of \(0.12\Msun{}\),
which corresponds to \(19\%\). Within the extended \Gaia{} mission the star passes 12 further background sources.
By combining the information of all astrometric microlensing events by LAWD 37 this result can improved slightly (see Fig.\,\ref{figure:diff_ideas_prox} bottom panel). 
We then expect a precision of \(0.10\,\Msun{}\) (\(16\%\)).

For 8 of the 22 lenses with multiple events the expected standard error is better than 50\%. The results of these events are given in Table\,\ref{table:multi}. For further three events the expected precision is between 50\% and 100\% (Figs.\,\ref{figure:diff_ideas_all}(g) to \ref{figure:diff_ideas_all}(i)).
Additionally to our three cases a more detailed selection of the used background sources can be done, however, this is only meaningful once the quality of the real data is known.

\phantom{ }
\footnotetext[9]{Spectro-Polarimetric High-contrast Exoplanet REsearch}
\footnotetext[10]{For an explanation of Violin plot see NIST: \url{https://www.itl.nist.gov/div898/software/dataplot/refman1/auxillar/violplot.htm}}
%

\section{Summary and Conclusion}
\label{chapter:conclusion}
In this work we showed that \Gaia{} can determine stellar masses for single stars using astrometric microlensing. For that purpose we simulated the individual \Gaia{}  measurements for 501 predicted events during the \Gaia{} era, using conservative cuts on the resolution and precision of Gaia.

In a similar study, \citet{2018MNRAS.476.2013R} showed that \Gaia{} might be able to measure the astrometric deflection caused by a stellar-mass black hole (\(M \approx 10 \Msun{}\)), based on results from a  photometric microlensing event detected by OGLE \citep{2016MNRAS.458.3012W}. Further, they claimed that for faint background sources (\(G> 17.5\,\mathrm{mag}\)) \Gaia{} might be able to detect the deflection of  black holes more massive than \(30 \Msun{}\). In the present paper, however, we consider bright lenses, which are also observed by Gaia. Hence, due to the additional measurements of the lens positions, we found that \Gaia{} can measure much smaller masses.

In this study we did not consider orbital motion. However, the orbital motion can be included in the fitting routine for the analysis of the real \Gaia{} measurements. \Gaia{} DR3 (expected for end of 2021) will include orbital parameters for a fraction of the contained stars. This information can be used to decide if orbital motion has to be considered or not.

We also assumed that  source and lens can only be resolved if both have individual readout windows. However, it might be possible to measure the separation in along-scan direction even 
from the blended measurement in one readout window. 
Due to the full width at half maximum of \(103\,\mathrm{mas}\) \citep{2016A&A...595A...3F} \Gaia{} might be able to resolve much closer lens-source pairs.
The astrometric microlensing signal of such measurements is stronger. Hence, the results of events with impact parameters smaller than the window size can be improved by a careful analysis of the data. 
Efforts in this direction are foreseen by the Gaia consortium for Gaia DR4 and DR5. 

Via a Monte Carlo approach we determined the expected precision of the mass determination and found that 
for 34 events a precision better than \(30\%\), and sometimes  down to \(5\%\) can be achieved.
By varying the input parameters we found that our results depend only weakly on selected input parameters. The scatter is of  the order of  \(6 \%\)  if the proper motion of the background star is known from \Gaia{} DR2 and of the order of \(10 \%\)  if the proper motion is unknown. Further, the dependency on the selected input mass is even weaker.

For \(17\) future events (closest approach after 2019.5), the \Gaia{} Data alone are not sufficient to derive a precise mass. 
For these events, it will be helpful to take further observations using, for example, HST, VLT or the Very Large Telescope Interferometer (VLTI).  Such two-dimensional measurements can easily be included in our fitting routine by adding two observations with perpendicular scanning directions. 
We showed that two additional highly accurate measurements can improve the results significantly, especially when the impact parameter of the event is smaller than \(1''\).
However, since the results depend on the resolution and precision of the additional observations, these properties should be implemented for such analyses, which is easily achievable. 
By doing so, our code can be a powerful tool to investigate different observation strategies.
The combination of Gaia data and additional information might also lead to better mass constraints for the two previously observed astrometric microlensing events of Stein 51b \citep{2017Sci...356.1046S}  and Proxima Centauri \citep{2018MNRAS.480..236Z}. For the latter, Gaia DR2 does not contain the background sources. But we are confident that Gaia has observed both background stars.
Finally, once the individual Gaia Measurements are published (DR4 or final data release), the code can be used to analyse the data, which will result in multiple well-measured masses of single stars. 
The code can also be used to fit the motion of multiple background sources simultaneously.
When combining these data, \Gaia{} can determine the mass of Proxima Centauri with a precision of  \(0.012\,\Msun{}\).

\begin{acknowledgements}
We gratefully thank the anonymous referee, whose suggestions greatly improved the paper.
This work has made use of results from the ESA space mission \Gaia{}, the data from which were 
processed by the  \Gaia{} Data Processing and Analysis Consortium (DPAC). Funding for the DPAC 
has been provided by national institutions, in particular the institutions participating in the  \Gaia{} 
Multilateral Agreement. The \Gaia{} mission website is:
\url{http://www.cosmos.esa.int/Gaia}. Two (U.\,B., J.\,K.) of the authors are members of the \Gaia{} Data Processing 
and Analysis Consortium (DPAC). 
This research has made use of the SIMBAD database,
operated at CDS, Strasbourg, France. This research made use of Astropy, a community-developed core Python package for Astronomy \citep{refId0}.
This research made use of matplotlib, a Python library for publication quality graphics \citep{Hunter:2007}. 
This research made use of SciPy \citep{jones_scipy_2001}. 
This research made use of NumPy \citep{van2011numpy}.

\end{acknowledgements}

\bibliographystyle{aa}

\renewcommand*{\arraystretch}{1.4}
\begin{table*}[]
\caption{Estimated uncertainties of mass measurements with astrometric microlensing with \Gaia{} for single events with an epoch of the closest approach during the nominal \Gaia{} mission. The table lists the name (Name-Lens) and \Gaia{} DR2 source ID (\(DR2\_ID\)-Lens) of the lens and the source ID of the background sources (\(DR2\_ID\)-Source). An asterisk indicates that  \Gaia{} DR2 provides only the position of the sources. Further, the table lists the epoch of the closest approach (\(T_{CA}\)) and the assumed mass of the lens (\(M_{in}\)). The expected precision for the use of  the data of the extended 10 years mission are given in \((\sigma\,M_{10})\), including the uncertainty due to the errors in the input parameters, and as percentage  \((\sigma\,M_{10}/M_{in})\). The expected precision for the use of the nominal 5 years mission is given in  \((\sigma\,M_{5})\) if it is below 100\% of the input mass. }
\label{table:single}
\tiny\begin{tabular}{l|rrrrr|rr|r|}
\# & Name-Lens & \(DR2\_ID\)-Lens & \(DR2\_ID\)-Source & \(T_{CA}\) & \(M_{in}\) & \(\sigma\,M_{10}\) & \(\sigma\,M_{10}/M_{in}\) & \(\sigma\,M_{5}  \) \\
  & & & & \(\mathrm{Jyear}\) & \(\mathrm{M_{\odot}}\) & \(\mathrm{M_{\odot}}\) & & \(\mathrm{M_{\odot}}\) \\
\hline
\(1\) & HD 22399 & \(488099359330834432\) & \(488099363630877824^{*}\) & \(2013.711\) & \(1.3\) & \(\pm0.60^{+0.07}_{-0.05}\)  & \(47\%\) &  \\
\(2\) & HD 177758 & \(4198685678421509376\) & \(4198685678400752128^{*}\) & \(2013.812\) & \(1.1\) & \(\pm0.41^{+0.04}_{-0.04}\)  & \(36\%\) &  \\
\(3\) & L 820-19 & \(5736464668224470400\) & \(5736464668223622784^{*}\) & \(2014.419\) & \(0.28\) & \(\pm0.058^{+0.004}_{-0.004}\)  & \(21\%\) & \(\pm0.11^{+0.01}_{-0.01}\) \\
\(4\) &  & \(2081388160068434048\) & \(2081388160059813120^{*}\) & \(2014.526\) & \(0.82\) & \(\pm0.39^{+0.05}_{-0.04}\)  & \(47\%\) &  \\
\(5\) &  & \(478978296199510912\) & \(478978296204261248\) & \(2014.692\) & \(0.65\) & \(\pm0.24^{+0.02}_{-0.02}\)  & \(36\%\) & \(\pm0.49^{+0.04}_{-0.04}\) \\
\rowcolor{lightGray}
\(6\) & G 123-61B & \(1543076475514008064\) & \(1543076471216523008^{*}\) & \(2014.763\) & \(0.26\) & \(\pm0.099^{+0.010}_{-0.009}\)  & \(37\%\) & \(\pm0.18^{+0.02}_{-0.02}\) \\
\rowcolor{lightGray}
\(7\) & L 601-78 & \(5600272625752039296\) & \(5600272629670698880^{*}\) & \(2014.783\) & \(0.21\) & \(\pm0.041^{+0.003}_{-0.004}\)  & \(19\%\) & \(\pm0.068^{+0.006}_{-0.005}\) \\
\rowcolor{lightGray}
\(8\) & UCAC3 27-74415 & \(6368299918479525632\) & \(6368299918477801728^{*}\) & \(2015.284\) & \(0.36\) & \(\pm0.047^{+0.005}_{-0.005}\)  & \(13\%\) & \(\pm0.11^{+0.01}_{-0.01}\) \\
\rowcolor{lightGray}
\(9\) & BD+00 5017 & \(2646280705713202816\) & \(2646280710008284416^{*}\) & \(2015.471\) & \(0.58\) & \(\pm0.18^{+0.02}_{-0.02}\)  & \(30\%\) & \(\pm0.44^{+0.04}_{-0.05}\) \\
\rowcolor{lightGray}
\(10\) & G 123-61A & \(1543076475509704192\) & \(1543076471216523008^{*}\) & \(2016.311\) & \(0.32\) & \(\pm0.079^{+0.007}_{-0.008}\)  & \(24\%\) & \(\pm0.14^{+0.02}_{-0.02}\) \\
\(11\) & EC 19249-7343 & \(6415630939116638464\) & \(6415630939119055872^{*}\) & \(2016.650\) & \(0.26\) & \(\pm0.099^{+0.008}_{-0.008}\)  & \(38\%\) &  \\
\(12\) &  & \(5312099874809857024\) & \(5312099870497937152\) & \(2016.731\) & \(0.07\) & \(\pm0.0088^{+0.0006}_{-0.0006}\)  & \(12\%\) & \(\pm0.0098^{+0.0010}_{-0.0007}\) \\
\(13\) & PM J08503-5848 & \(5302618648583292800\) & \(5302618648591015808\) & \(2017.204\) & \(0.65\) & \(\pm0.17^{+0.02}_{-0.02}\)  & \(26\%\) & \(\pm0.29^{+0.03}_{-0.03}\) \\
\(14\) &  & \(5334619419176460928\) & \(5334619414818244992\) & \(2017.258\) & \(0.65\) & \(\pm0.23^{+0.02}_{-0.02}\)  & \(35\%\) & \(\pm0.43^{+0.04}_{-0.03}\) \\
\(15\) & Proxima Cen & \(5853498713160606720\) & \(5853498708818460032\) & \(2017.392\) & \(0.12\) & \(\pm0.034^{+0.003}_{-0.003}\)  & \(29\%\) & \(\pm0.082^{+0.006}_{-0.006}\) \\
\rowcolor{lightGray}
\(16\) & Innes' star & \(5339892367683264384\) & \(5339892367683265408\) & \(2017.693\) & \(0.33\) & \(\pm0.16^{+0.02}_{-0.01}\)  & \(48\%\) & \(\pm0.22^{+0.02}_{-0.02}\) \\
\rowcolor{lightGray}
\(17\) & L 51-47 & \(4687511776265158400\) & \(4687511780573305984\) & \(2018.069\) & \(0.28\) & \(\pm0.035^{+0.003}_{-0.003}\)  & \(12\%\) & \(\pm0.046^{+0.004}_{-0.004}\) \\
\rowcolor{lightGray}
\(18\) &  & \(4970215770740383616\) & \(4970215770743066240\) & \(2018.098\) & \(0.65\) & \(\pm0.076^{+0.008}_{-0.005}\)  & \(12\%\) & \(\pm0.21^{+0.02}_{-0.02}\) \\
\rowcolor{lightGray}
\(19\) & BD-06 855 & \(3202470247468181632\) & \(3202470247468181760^{*}\) & \(2018.106\) & \(0.8\) & \(\pm0.38^{+0.05}_{-0.05}\)  & \(46\%\) & \(\pm0.72^{+0.19}_{-0.08}\) \\
\rowcolor{lightGray}
\(20\) & G 217-32 & \(429297924157113856\) & \(429297859741477888^{*}\) & \(2018.134\) & \(0.23\) & \(\pm0.018^{+0.002}_{-0.002}\)  & \(7.5\%\) & \(\pm0.040^{+0.003}_{-0.004}\) \\
\(21\) &  & \(5865259639247544448\) & \(5865259639247544064^{*}\) & \(2018.142\) & \(0.8\) & \(\pm0.053^{+0.008}_{-0.006}\)  & \(6.6\%\) & \(\pm0.12^{+0.02}_{-0.02}\) \\
\(22\) & HD 146868 & \(1625058605098521600\) & \(1625058605097111168^{*}\) & \(2018.183\) & \(0.92\) & \(\pm0.060^{+0.005}_{-0.005}\)  & \(6.5\%\) & \(\pm0.14^{+0.02}_{-0.02}\) \\
\(23\) & Ross 733 & \(4516199240734836608\) & \(4516199313402714368\) & \(2018.282\) & \(0.43\) & \(\pm0.067^{+0.005}_{-0.006}\)  & \(15\%\) & \(\pm0.13^{+0.01}_{-0.02}\) \\
\(24\) & LP 859-51 & \(6213824650812054528\) & \(6213824650808938880^{*}\) & \(2018.359\) & \(0.43\) & \(\pm0.19^{+0.03}_{-0.03}\)  & \(43\%\) &  \\
\(25\) & L 230-188 & \(4780100658292046592\) & \(4780100653995447552\) & \(2018.450\) & \(0.15\) & \(\pm0.073^{+0.007}_{-0.004}\)  & \(49\%\) & \(\pm0.11^{+0.01}_{-0.01}\) \\
\rowcolor{lightGray}
\(26\) & HD 149192 & \(5930568598406530048\) & \(5930568568425533440^{*}\) & \(2018.718\) & \(0.67\) & \(\pm0.043^{+0.007}_{-0.004}\)  & \(6.3\%\) & \(\pm0.12^{+0.01}_{-0.01}\) \\
\rowcolor{lightGray}
\(27\) & HD 85228 & \(5309386791195469824\) & \(5309386795502307968^{*}\) & \(2018.751\) & \(0.82\) & \(\pm0.043^{+0.004}_{-0.004}\)  & \(5.1\%\) & \(\pm0.65^{+0.05}_{-0.06}\) \\
\rowcolor{lightGray}
\(28\) & HD 155918 & \(5801950515627094400\) & \(5801950515623081728^{*}\) & \(2018.773\) & \(1\) & \(\pm0.13^{+0.02}_{-0.02}\)  & \(13\%\) & \(\pm0.35^{+0.03}_{-0.03}\) \\
\rowcolor{lightGray}
\(29\) & HD 77006 & \(1015799283499485440\) & \(1015799283498355584^{*}\) & \(2018.796\) & \(1\) & \(\pm0.42^{+0.05}_{-0.05}\)  & \(41\%\) &  \\
\rowcolor{lightGray}
\(30\) & Proxima Cen & \(5853498713160606720\) & \(5853498713181091840\) & \(2018.819\) & \(0.12\) & \(\pm0.020^{+0.002}_{-0.002}\)  & \(16\%\) & \(\pm0.052^{+0.005}_{-0.005}\) \\
\(31\) & HD 2404 & \(2315857227976341504\) & \(2315857227975556736^{*}\) & \(2019.045\) & \(0.84\) & \(\pm0.19^{+0.02}_{-0.02}\)  & \(22\%\) &  \\
\(32\) & LP 350-66 & \(2790883634570755968\) & \(2790883634570196608^{*}\) & \(2019.299\) & \(0.32\) & \(\pm0.15^{+0.02}_{-0.02}\)  & \(46\%\) &  \\
\(33\) & HD 110833 & \(1568219729458240128\) & \(1568219729456499584^{*}\) & \(2019.360\) & \(0.78\) & \(\pm0.056^{+0.005}_{-0.004}\)  & \(7.2\%\) &  \\
\hline
\end{tabular}
\end{table*}

\begin{table*}[]
\caption{Estimated uncertainties of mass measurements with astrometric microlensing with \Gaia{} for single events with an epoch of the closest approach after 2019.5. The table lists the name (Name-Lens) and \Gaia{} DR2 source ID (\(DR2\_ID\)-Lens) of the lens and the source ID of the background sources (\(DR2\_ID\)-Source). An asterisk indicates that  \Gaia{} DR2 provides only the position of the sources. Further, the table lists the epoch of the closest approach (\(T_{CA}\)) and the assumed mass of the lens (\(M_{in}\)). The expected precision for the use of  the data of the extended 10 years mission are given in \((\sigma\,M_{10})\), including the uncertainty due to the errors in the input parameters,  and as percentage  \((\sigma\,M_{10}/M_{in})\). The expected relative precision while including two external 2D observations is given in \((\sigma\,M_{obs}/M_{in})\)}
\label{table:single2}
\tiny\begin{tabular}{l|rrrrr|rr|r|}
\# & Name-Lens & \(DR2\_ID\)-Lens & \(DR2\_ID\)-Source & \(T_{CA}\) & \(M_{in}\) & \(\sigma\,M_{10}\) & \(\sigma\,M_{10}/M_{in}\) & \(\sigma\,M_{obs}/M_{in}\)\\
  & & & & \(\mathrm{Jyear}\) & \(\mathrm{M_{\odot}}\) & \(\mathrm{M_{\odot}}\) & & \\
\hline
\(34\) & L 702-43 & \(4116840541184279296\) & \(4116840536790875904\) & \(2019.673\) & \(0.22\) & \(\pm0.085^{+0.006}_{-0.006}\)  & \(39\%\) & \(27\%\)\\
\(35\) & G 251-35 & \(1136512191212093440\) & \(1136512191210614272^{*}\) & \(2019.856\) & \(0.64\) & \(\pm0.26^{+0.02}_{-0.03}\)  & \(39\%\) & \(39\%\)\\
\(36\) & G 245-47A & \(546488928621555328\) & \(546488928619704320\) & \(2019.858\) & \(0.27\) & \(\pm0.075^{+0.006}_{-0.006}\)  & \(28\%\) & \(26\%\)\\
\(37\) & LAWD 37 & \(5332606522595645952\) & \(5332606350796955904\) & \(2019.865\) & \(0.65\) & \(\pm0.13^{+0.02}_{-0.01}\)  & \(19\%\) & \(12\%\)\\
\(38\) & LSPM J2129+4720 & \(1978296747258230912\) & \(1978296747268742784^{*}\) & \(2019.960\) & \(0.34\) & \(\pm0.13^{+0.02}_{-0.02}\)  & \(38\%\) & \(9.2\%\)\\
\rowcolor{lightGray}
\(39\) &  & \(5788178166117584640\) & \(5788178170416392192^{*}\) & \(2020.126\) & \(0.54\) & \(\pm0.12^{+0.02}_{-0.02}\)  & \(22\%\) & \(6.5\%\)\\
\rowcolor{lightGray}
\(40\) & G 16-29 & \(4451575895403432064\) & \(4451575895400387968^{*}\) & \(2020.214\) & \(0.51\) & \(\pm0.19^{+0.05}_{-0.03}\)  & \(37\%\) & \(17\%\)\\
\rowcolor{lightGray}
\(41\) & HD 66553 & \(654826970401335296\) & \(654826970399770368^{*}\) & \(2020.309\) & \(0.87\) & \(\pm0.11^{+0.02}_{-0.02}\)  & \(12\%\) & \(9\%\)\\
\rowcolor{lightGray}
\(42\) & HD 120065 & \(1251328585567327744\) & \(1251328589861875840^{*}\) & \(2020.344\) & \(1.2\) & \(\pm0.54^{+0.09}_{-0.06}\)  & \(44\%\) & \(42\%\)\\
\rowcolor{lightGray}
\(43\) & HD 78663 & \(3842095911266162432\) & \(3842095915561269888^{*}\) & \(2020.356\) & \(0.8\) & \(\pm0.28^{+0.05}_{-0.04}\)  & \(34\%\) & \(32\%\)\\
\(44\) & HD 124584 & \(5849427049801267200\) & \(5849427114177683584^{*}\) & \(2020.742\) & \(1.2\) & \(\pm0.47^{+0.06}_{-0.04}\)  & \(39\%\) & \(39\%\)\\
\(45\) & Proxima Cen & \(5853498713160606720\) & \(5853498713181092224\) & \(2020.823\) & \(0.12\) & \(\pm0.023^{+0.002}_{-0.002}\)  & \(19\%\) & \(19\%\)\\
\(46\) & HD 222506 & \(2390377345808152832\) & \(2390377350103204096^{*}\) & \(2021.017\) & \(0.86\) & \(\pm0.20^{+0.03}_{-0.02}\)  & \(23\%\) & \(9.8\%\)\\
\(47\) &  & \(3670594366739201664\) & \(3670594366739201536^{*}\) & \(2021.115\) & \(0.65\) & \(\pm0.13^{+0.02}_{-0.02}\)  & \(18\%\) & \(5.8\%\)\\
\(48\) & L 143-23 & \(5254061535097566848\) & \(5254061535097574016\) & \(2021.188\) & \(0.12\) & \(\pm0.050^{+0.004}_{-0.004}\)  & \(43\%\) & \(21\%\)\\
\rowcolor{lightGray}
\(49\) & HD 44573 & \(2937651222655480832\) & \(2937651222651937408^{*}\) & \(2021.226\) & \(0.77\) & \(\pm0.18^{+0.02}_{-0.02}\)  & \(22\%\) & \(22\%\)\\
\rowcolor{lightGray}
\(50\) & CD-32 12693 & \(5979367986779538432\) & \(5979367982463635840\) & \(2021.336\) & \(0.43\) & \(\pm0.13^{+0.01}_{-0.01}\)  & \(28\%\) & \(25\%\)\\
\rowcolor{lightGray}
\(51\) & 75 Cnc & \(689004018040211072\) & \(689004018038546560^{*}\) & \(2021.378\) & \(1.4\) & \(\pm0.11^{+0.01}_{-0.01}\)  & \(7.6\%\) & \(5.4\%\)\\
\rowcolor{lightGray}
\(52\) & OGLE SMC115.5 319 & \(4687445500635789184\) & \(4687445599404851456\) & \(2021.500\) & \(0.65\) & \(\pm0.13^{+0.01}_{-0.01}\)  & \(20\%\) & \(12\%\)\\
\rowcolor{lightGray}
\(53\) & HD 197484 & \(6677000246203170944\) & \(6677000246201701120^{*}\) & \(2021.579\) & \(0.99\) & \(\pm0.26^{+0.06}_{-0.04}\)  & \(26\%\) & \(12\%\)\\
\(54\) & LAWD 37 & \(5332606522595645952\) & \(5332606350771972352\) & \(2022.226\) & \(0.65\) & \(\pm0.31^{+0.03}_{-0.03}\)  & \(47\%\) & \(47\%\)\\
\(55\) & BD+43 4138 & \(1962597885872344704\) & \(1962597885867565312^{*}\) & \(2022.663\) & \(0.96\) & \(\pm0.48^{+0.07}_{-0.05}\)  & \(49\%\) & \(49\%\)\\
\(56\) & L 100-115 & \(5243594081269535872\) & \(5243594081263121792\) & \(2022.847\) & \(0.17\) & \(\pm0.054^{+0.005}_{-0.004}\)  & \(31\%\) & \(27\%\)\\
\(57\) & G 100-35B & \(3398414352092062720\) & \(3398414347798489472\) & \(2022.859\) & \(0.25\) & \(\pm0.093^{+0.007}_{-0.007}\)  & \(37\%\) & \(33\%\)\\
\(58\) & Proxima Cen & \(5853498713160606720\) & \(5853498713180840704\) & \(2023.349\) & \(0.12\) & \(\pm0.047^{+0.005}_{-0.004}\)  & \(40\%\) & \(40\%\)\\
\rowcolor{lightGray}
\(59\) &  & \(5340884333294888064\) & \(5340884328994222208\) & \(2023.559\) & \(0.65\) & \(\pm0.32^{+0.03}_{-0.03}\)  & \(49\%\) & \(48\%\)\\
\rowcolor{lightGray}
\(60\) & HD 18757 & \(466294295706341760\) & \(466294295706435712\) & \(2023.762\) & \(1\) & \(\pm0.36^{+0.03}_{-0.03}\)  & \(35\%\) & \(34\%\)\\
\rowcolor{lightGray}
\(61\) & Proxima Cen & \(5853498713160606720\) & \(5853498713180846592\) & \(2023.849\) & \(0.12\) & \(\pm0.034^{+0.002}_{-0.003}\)  & \(28\%\) & \(27\%\)\\
\rowcolor{lightGray}
\(62\) & OGLE LMC162.5 41235 & \(4657982643495556608\) & \(4657982639152973184\) & \(2024.283\) & \(0.65\) & \(\pm0.23^{+0.02}_{-0.02}\)  & \(35\%\) & \(30\%\)\\
\rowcolor{lightGray}
\(63\) & L 31-84 & \(4623882630333283328\) & \(4623882630333283968\) & \(2024.593\) & \(0.35\) & \(\pm0.088^{+0.007}_{-0.007}\)  & \(25\%\) & \(12\%\)\\
\(64\) & 61 Cyg B & \(1872046574983497216\) & \(1872046605038072448\) & \(2024.661\) & \(0.55\) & \(\pm0.14^{+0.01}_{-0.01}\)  & \(24\%\) & \(11\%\)\\
\(65\) & HD 207450 & \(6585158207436506368\) & \(6585158211732350592^{*}\) & \(2025.805\) & \(1.2\) & \(\pm0.44^{+0.06}_{-0.09}\)  & \(35\%\) & \(3.5\%\)\\
\hline
\end{tabular}
\end{table*}
\renewcommand*{\arraystretch}{1.4}
\begin{sidewaystable*}[]
\caption{ Estimated uncertainties of mass measurements with astrometric microlensing with \Gaia{} for multiple background sources. The table lists the name (Name-Lens) and \Gaia{} DR2 source ID  \((DR2\_ID)\) of the lens and DR2 source IDs of the background sources. The background sources are grouped into sources where \Gaia{} DR2 does only provides only the position (2-parameter), sources with a full 5-parameter solution (5-parameter), and sources with a 5-parameter solution in combination with an expected precision in along-scan direction better than of \(0.5\,\mathrm{mas}\) (sigma). The assumed mass of the lens is listed in \(M_{in}\), and the expected precisions of the mass determination for the three cases are given in \(\sigma\,M_{x}\), including the uncertainty due to the errors in the input parameters, as well as the percentage in \(\sigma\,M_{x}/M_{in}\), where x indicates the use of all background stars (all), background stars with a 5-parameters solution (5-par.) and background stars with expected precision in along-scan direction better than of \(0.5\,\mathrm{mas}\) (sig.).}
\label{table:multi}
\tiny\begin{tabular}{l|r|rrrr|rr|rr|rr|}
\# & Name-Lens & 
\(DR2\_ID\)-Source & \(DR2\_ID\)-Source & \(DR2\_ID\)-Source
 & \(M_{\mathrm{in}}\) & \(\sigma\,M_{\mathrm{all}}\) & \(\sigma\,M_{\mathrm{all}}/M_{\mathrm{in}}\) & \(\sigma\,M_{\mathrm{5-par}}  \) & \(\sigma\,M_{\mathrm{5-par.}}/M_{in}\)  & \(\sigma\,M_{\mathrm{sig.}}  \) & \(\sigma\,M_{\mathrm{sig.}}/M_{\mathrm{in}}\)  \\
 & \(DR2\_ID\)-Lens & (2-parameter) & (5-parameter)& (sigma) &  \(\Msun{}\) & \(\Msun{}\) & \(\%\)& \(\Msun{}\) & \(\%\)& \(\Msun{}\) & \(\%\) \\
\hline
1 &  Proxima Cen  & \(5853498713161902848\) & \(5853498713162466688\) & \(5853498713181091840\)  & \(0.12\)  &\(\pm0.012^{+0.001}_{-0.001}\) & \(10\%\) &\(\pm0.012^{+0.001}_{-0.001}\) &\(10\%\) &\(\pm0.012^{+0.001}_{-0.001}\) & \(10\%\)  \\
 &\(5853498713160606720\)& \(5853498713161549312\) & \(5853498713161890304\) & \(5853498640098981888\)  & & & & & & & \\
 && \(\) & \(5853498713160604800\) & \(5853498713180846720\)  & & & & & & & \\
 && \(\) & \(5853498644442433280\) & \(5853498713160651648\)  & & & & & & & \\
 && \(\) & \(5853498644461348864\) & \(5853498713180846592\)  & & & & & & & \\
 && \(\) & \(5853498713160648448\) & \(5853498713181092224\)  & & & & & & & \\
 && \(\) & \(5853498713160633088\) & \(5853498713180840704\)  & & & & & & & \\
 && \(\) & \(5853498713161549440\) & \(5853498708818460032\)  & & & & & & & \\
\hline
\rowcolor{lightGray}
2 &  \(5312099874809857024\)  & \(\) & \(\) & \(5312099870497937152\)  & \(0.07\)  & & & & &\(\pm0.0088^{+0.0007}_{-0.0006}\) & \(12\%\)  \\
\rowcolor{lightGray}
 && \(\) & \(\) & \(5312099874795679488\)  & & & & & & & \\
\hline
3 &  Ross 733  & \(4516199313384903936\) & \(\) & \(4516199313402714368\)  & \(0.43\)  &\(\pm0.065^{+0.004}_{-0.005}\) & \(15\%\) & & &\(\pm0.066^{+0.005}_{-0.006}\) & \(15\%\)  \\
 &\(4516199240734836608\)& \(\) & \(\) & \(\)  & & & & & & & \\
\hline
\rowcolor{lightGray}
4 &  LAWD 37  & \(\) & \(5332606346480229376\) & \(5332606522595645824\)  & \(0.65\)  & & &\(\pm0.11^{+0.01}_{-0.01}\) &\(16\%\) &\(\pm0.24^{+0.02}_{-0.02}\) & \(36\%\)  \\
\rowcolor{lightGray}
 &\(5332606522595645952\)& \(\) & \(5332606350774082944\) & \(5332606350771972352\)  & & & & & & & \\
\rowcolor{lightGray}
 && \(\) & \(5332606350774081280\) & \(\)  & & & & & & & \\
\rowcolor{lightGray}
 && \(\) & \(5332606522574285952\) & \(\)  & & & & & & & \\
\rowcolor{lightGray}
 && \(\) & \(5332606350796954240\) & \(\)  & & & & & & & \\
\rowcolor{lightGray}
 && \(\) & \(5332606350796955904\) & \(\)  & & & & & & & \\
\rowcolor{lightGray}
 && \(\) & \(5332606316412235008\) & \(\)  & & & & & & & \\
\rowcolor{lightGray}
 && \(\) & \(5332606522573342592\) & \(\)  & & & & & & & \\
\rowcolor{lightGray}
 && \(\) & \(5332606518268141952\) & \(\)  & & & & & & & \\
\rowcolor{lightGray}
 && \(\) & \(5332606522572755072\) & \(\)  & & & & & & & \\
\hline
5 &  61 Cyg B  & \(1872046609337556608\) & \(\) & \(1872046605038072448\)  & \(0.55\)  &\(\pm0.14^{+0.01}_{-0.01}\) & \(24\%\) & & &\(\pm0.14^{+0.01}_{-0.02}\) & \(24\%\)  \\
 &\(1872046574983497216\)& \(1872046609336784256\) & \(\) & \(\)  & & & & & & & \\
\hline
\rowcolor{lightGray}
6 &  61 Cyg A  & \(\) & \(1872046609343242368\) & \(1872047330897752832\)  & \(0.63\)  & & &\(\pm0.21^{+0.02}_{-0.02}\) &\(32\%\) &\(\pm0.33^{+0.03}_{-0.03}\) & \(53\%\)  \\
\rowcolor{lightGray}
 &\(1872046574983507456\)& \(\) & \(1872046609335739648\) & \(\)  & & & & & & & \\
\rowcolor{lightGray}
 && \(\) & \(1872046605041836928\) & \(\)  & & & & & & & \\
\rowcolor{lightGray}
 && \(\) & \(1872046609335748736\) & \(\)  & & & & & & & \\
\hline
7 &  L 143-23  & \(\) & \(5254061535052907008\) & \(5254061535097574016\)  & \(0.12\)  & & &\(\pm0.043^{+0.003}_{-0.003}\) &\(37\%\) &\(\pm0.051^{+0.004}_{-0.004}\) & \(44\%\)  \\
 &\(5254061535097566848\)& \(\) & \(5254061535052928128\) & \(\)  & & & & & & & \\
\hline
\rowcolor{lightGray}
8 &  Innes' star  & \(\) & \(5339892367648234240\) & \(5339892367683265408\)  & \(0.33\)  & & &\(\pm0.16^{+0.01}_{-0.01}\) &\(47\%\) &\(\pm0.16^{+0.02}_{-0.02}\) & \(48\%\)  \\
\rowcolor{lightGray}
 &\(5339892367683264384\)& \(\) & \(\) & \(\)  & & & & & & & \\
\hline
\end{tabular}
\end{sidewaystable*}

\end{document}